\documentclass[
aps,
prapplied,
twocolumn, 
longbibliography, 
superscriptaddress,
floatfix
]{revtex4-2}

\usepackage{amsmath,amssymb}
\usepackage{graphicx}
\usepackage[percent]{overpic}
\usepackage{hyperref}
\usepackage{caption} 
\usepackage{subcaption}

\begin{document}

\title{Enhancement of spin current in Fe$_{85}$Co$_{15}$/Ni$_{80}$Fe$_{20}$ bilayers via interlayer ferromagnetic coupling}

\author{A. A. Pérez Martínez}
\email{aaperezmtnez92@gmail.com}
\affiliation{Instituto de Nanociencia y Nanotecnología (CNEA-CONICET), Nodo Bariloche, Av. Bustillo 9500, (8400) Bariloche, Río Negro, Argentina}
\affiliation{Departamento de Magnetismo y Materiales Magnéticos, Gerencia de Física, Centro Atómico Bariloche, Av. E. Bustillo 9500, San Carlos de Bariloche, (R8402AGP), Río Negro, Argentina}
\affiliation{Instituto Balseiro, Universidad Nacional de Cuyo, Av. Bustillo 9500, (8400) Bariloche, Río Negro, Argentina}

\author{D. Velázquez Rodríguez}
\affiliation{Instituto de Nanociencia y Nanotecnología (CNEA-CONICET), Nodo Bariloche, Av. Bustillo 9500, (8400) Bariloche, Río Negro, Argentina}
\affiliation{Departamento de Magnetismo y Materiales Magnéticos, Gerencia de Física, Centro Atómico Bariloche, Av. E. Bustillo 9500, San Carlos de Bariloche, (R8402AGP), Río Negro, Argentina}

\author{D. Goijman}
\affiliation{Instituto de Nanociencia y Nanotecnología (CNEA-CONICET), Nodo Bariloche, Av. Bustillo 9500, (8400) Bariloche, Río Negro, Argentina}
\affiliation{Departamento de Magnetismo y Materiales Magnéticos, Gerencia de Física, Centro Atómico Bariloche, Av. E. Bustillo 9500, San Carlos de Bariloche, (R8402AGP), Río Negro, Argentina}

\author{T. Torres}
\affiliation{Instituto de Nanociencia y Nanotecnología (CNEA-CONICET), Nodo Bariloche, Av. Bustillo 9500, (8400) Bariloche, Río Negro, Argentina}
\affiliation{Departamento de Magnetismo y Materiales Magnéticos, Gerencia de Física, Centro Atómico Bariloche, Av. E. Bustillo 9500, San Carlos de Bariloche, (R8402AGP), Río Negro, Argentina}

\author{M. H. Aguirre}
\affiliation{Instituto de Nanociencia y Materiales de Aragón (INMA-CSIC) and Dpto. de Física de la Materia Condensada, Universidad de Zaragoza, Spain.}

\author{J. Gómez}
\affiliation{Instituto de Nanociencia y Nanotecnología (CNEA-CONICET), Nodo Bariloche, Av. Bustillo 9500, (8400) Bariloche, Río Negro, Argentina}
\affiliation{Departamento de Magnetismo y Materiales Magnéticos, Gerencia de Física, Centro Atómico Bariloche, Av. E. Bustillo 9500, San Carlos de Bariloche, (R8402AGP), Río Negro, Argentina}

\author{A. Butera}
\affiliation{Instituto de Nanociencia y Nanotecnología (CNEA-CONICET), Nodo Bariloche, Av. Bustillo 9500, (8400) Bariloche, Río Negro, Argentina}
\affiliation{Departamento de Magnetismo y Materiales Magnéticos, Gerencia de Física, Centro Atómico Bariloche, Av. E. Bustillo 9500, San Carlos de Bariloche, (R8402AGP), Río Negro, Argentina}
\affiliation{Instituto Balseiro, Universidad Nacional de Cuyo, Av. Bustillo 9500, (8400) Bariloche, Río Negro, Argentina}

\author{E. De Biasi}
\affiliation{Instituto de Nanociencia y Nanotecnología (CNEA-CONICET), Nodo Bariloche, Av. Bustillo 9500, (8400) Bariloche, Río Negro, Argentina}
\affiliation{Departamento de Magnetismo y Materiales Magnéticos, Gerencia de Física, Centro Atómico Bariloche, Av. E. Bustillo 9500, San Carlos de Bariloche, (R8402AGP), Río Negro, Argentina}
\affiliation{Instituto Balseiro, Universidad Nacional de Cuyo, Av. Bustillo 9500, (8400) Bariloche, Río Negro, Argentina}

\author{J. Milano}
\affiliation{Instituto de Nanociencia y Nanotecnología (CNEA-CONICET), Nodo Bariloche, Av. Bustillo 9500, (8400) Bariloche, Río Negro, Argentina}
\affiliation{Departamento de Magnetismo y Materiales Magnéticos, Gerencia de Física, Centro Atómico Bariloche, Av. E. Bustillo 9500, San Carlos de Bariloche, (R8402AGP), Río Negro, Argentina}
\affiliation{Instituto Balseiro, Universidad Nacional de Cuyo, Av. Bustillo 9500, (8400) Bariloche, Río Negro, Argentina}

\begin{abstract}

We present a detailed study on how the strength of the interlayer magnetic coupling on Fe$_{85}$Co$_{15}$/Ni$_{80}$Fe$_{20}$ bilayers modifies the spin wave behavior of this system. A series of Fe$_{85}$Co$_{15}$/Ni$_{80}$Fe$_{20}$ bilayers deposited on MgO[100] substrates were grown by magnetron sputtering. Magnetic characterization of the samples was performed using a vibrating sample magnetometer and magneto-optical Kerr effect. The in-plane hysteresis loops reveal a cubic magnetic anisotropy of magnetocrystalline origin, with easy and hard axis along the [100] and [110] Fe-Co crystallographic directions, respectively. Ferromagnetic resonance measurements were performed to analyze the in-plane angular dependence of the resonance field, and also the resonance field at several frequencies was determined along the hard axis. By using a bilayer model in the frame of the Landau-Lifshitz-Gilbert magnetization equation of motion, the magnetization precession components were calculated, as well as the dependence of precession area on the Fe-Co layer thickness and the ferromagnetic interlayer coupling. We observe a maximum in the area of the ellipsoid generated by the magnetization precession of the permalloy layer at a certain exchange constant, showing that this effect could be used to maximize the injected spin currents, which could be tuned by changing the interlayer exchange constant in bilayer systems, the saturation magnetization of the materials, or the excitation frequency.

\end{abstract}

\maketitle

\section{Introduction}
The main challenges in modern electronics are miniaturization, processing speed, and energy consumption \cite{puebla2020spintronic}. Regarding the latter, spintronics has emerged as a promising solution. In particular, pure spin currents can be used to transport information\cite{hoffmann2007pure}, which could lead to a considerable reduction in losses due to Joule effect. 

Among the most studied mechanisms for generation and detection of pure spin currents are Spin Pumping (SP), Spin Hall Effect (SHE) and the inverse Spin Hall Effect (ISHE)\cite{tserkovnyak2002enhanced, ando2008angular, saitoh2006conversion, valenzuela2006direct}. In the SP effect, the magnetization of a ferromagnet (FM) is driven to the the ferromagnetic resonance (FMR) condition by a microwave field, generating a pure spin current in an adjacent normal metal (NM). In this scenario, magnetization precession acts as a spin pump which transfers angular momentum from the ferromagnet to the normal metal. 

Most studies have focused on FM/NM systems to investigate Spin Pumping. In Ref. \cite{taniguchi2008spin} the authors developed the theoretical framework for spin pumping in ferromagnetic multilayers. In these systems, it becomes essential to consider the interlayer exchange energy and its influence on ferromagnetic resonance.  Several studies have examined exchange-coupled ferromagnetic multilayers \cite{layadi1990ferromagnetic,wigen1992ferromagnetic, zhang1994angular}, analyzing how dispersion relations, magnetization components, and mode intensities are affected.

The amplitude of spin current is proportional to the area of magnetization precession trajectory; a larger precession area results in a higher injected spin current \cite{ando2011inverse}. In this context, Fe–Co alloys emerge as promising candidates for the ferromagnetic layer. For instance, first-principles calculations reported in Ref. \cite{mankovsky2013first} predicts a minimum intrinsic Gilbert damping of $\alpha_\text{int}$$\thicksim$0.0005 for Fe–Co alloys with Co concentrations between 10~\% and 20~\%. Consistently, experimental results in Ref. \cite{rodriguez2024intrinsic} provide evidence of a minimum Gilbert damping at a Co concentration of approximately 15~\%.

In this paper, we present a study of a bilayer system composed of two ferromagnetic layers grown on a MgO [100] substrate, Fe$_{85}$Co$_{15}$ (Fe-Co) and Permalloy (Py), where the Fe-Co thickness is varied. We investigated the effect of the interface exchange energy on the FMR spectra as well as the influence of Fe-Co thickness, using an exchange-coupled bilayer model, and analyzed how these parameters modify the magnetization components and the spin current intensity generated by magnetization precession.

\section{Experimental}
\subsection{Growth and structural characterization}
\setlength{\parindent}{20pt}
\indent We deposited a series of MgO//Fe$_{85}$Co$_{15}$/Py bilayers on MgO [100] substrates using magnetron sputtering techniques. The series was composed of five bilayers in which the thickness of Fe-Co layer was varied from 5~nm to 25~nm with a step of 5~nm. The Py layer thickness was kept fixed at 5~nm. The Fe-Co layer was deposited using an Ar pressure of 1.8 mTorr and a power density of 3.97 W/$\text{cm}^2$, which gives a deposition rate of 0.07~nm/s. The Py layer was grown at an Ar pressure of 2.6~mTorr and a power density of 5.26~W/$\text{cm}^2$, which originates a rate of 0.13~nm/s. To check the magnetic behavior for a large thickness of FeCo and Py, two extra samples were grown, the first one Fe-Co(50~nm)/Py(3~nm), and the second Fe-Co(3~nm)/Py(50~nm), keeping the same layer sequence. All depositions were performed at room temperature.

The thicknesses were measured by X-Ray Reflectometry (XRR) and High Resolution Transmission Electron Microscopy (HRTEM) and the values found were similar to the nominal ones (see Table~\ref{table:ticknesses}). In addition, by using these techniques, we can appreciate the quality of the samples, as explained below. HRTEM analysis was not performed on the FeCo10 and FeCo20 samples.

\begin{table}[h!]
\centering
\begin{tabular}{c c c c c}
\hline
\hline
  & Fe-Co/Py(XRR)(nm) & Fe-Co/Py(HRTEM)(nm) \\ 
 \hline
 FeCo5 & 4.9(0.1)/5.5(0.1) & 4.9(0.2)/5.6(0.2) \\ 
 FeCo10 & 9.6(0.4)/4.2(0.3) &   \\ 
 FeCo15 & 13.6(0.4)/5.5(0.1) & 14.4(0.2)/5.6(0.2)  \\ 
 FeCo20 & 19.4(0.6)/5.4(0.2) &  \\ 
 FeCo25 & 25.7(0.6)/5.4(0.2) & 25.1(0.2)/5.6(0.2)  \\ 
\hline
\hline
\end{tabular}

\caption {Thicknesses measured by XRR and TEM. The number in the first column indicates the film thickness.}
\label{table:ticknesses}
\end{table}

\begin{figure}[h!]
   \centering
    \includegraphics[width=0.45\textwidth]{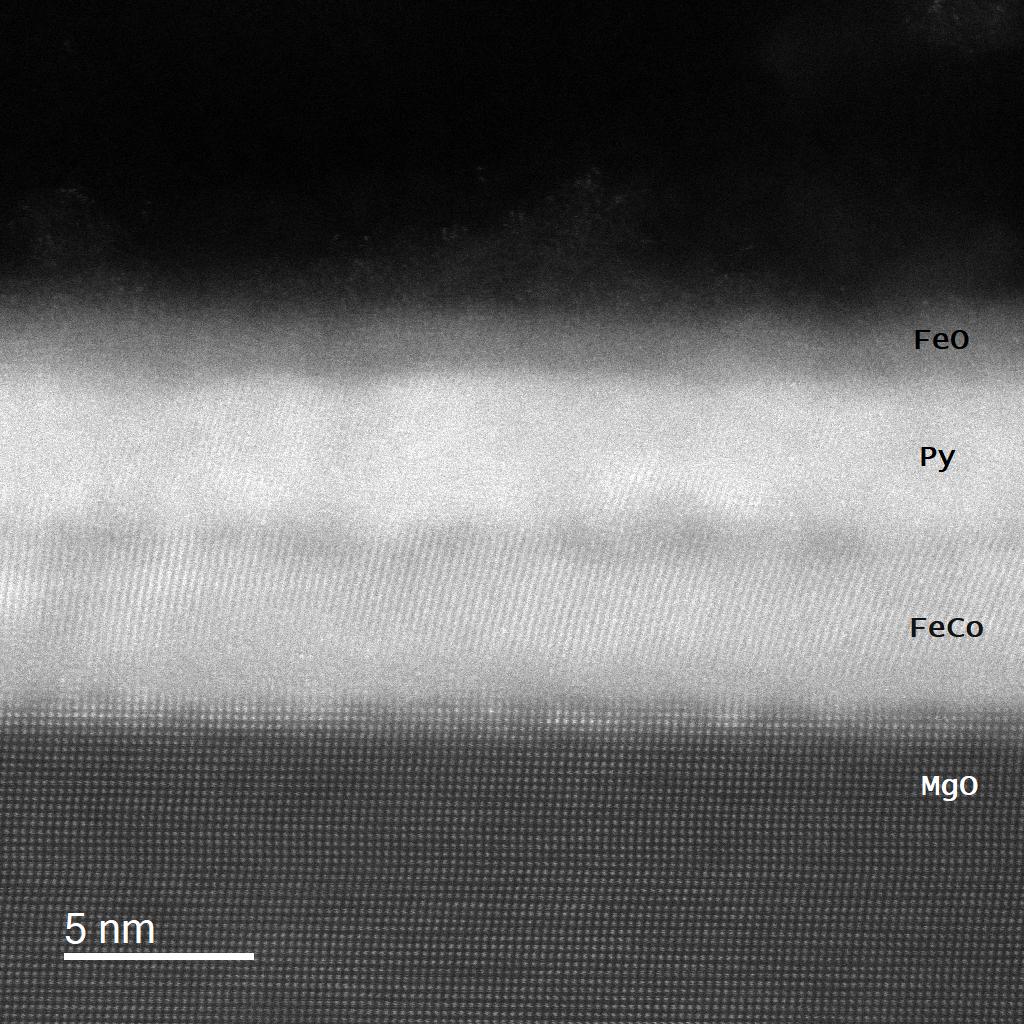}
    \caption{HRTEM image for the FeCo5 sample.}
    \label{TEM}
\end{figure}

In Fig.~\ref{TEM} one of the HRTEM images of the FeCo5nm/Py sample is shown. In the image we can observe that Fe-Co layer grows epitaxially on MgO but rotated by $45^\circ$, Fe-Co[001]~||~MgO[001] and Fe-Co[110]~||~MgO[100], as reported in Ref. \cite{rodriguez2024intrinsic}. The Py layer also follows the same epitaxy but with small amorphous zones. Also, a thin FeO capping layer ($\thicksim$ 1~nm) is formed on top of the unprotected Py film. The  film concentrations were corroborated by Electron Energy Loss Spectroscopy~(EELS), obtaining good agreement between the nominal values expected from the sputtering target and those measured in the deposited bilayer. In Ref. \cite{rodriguez2024intrinsic}, X-ray diffraction studies indicated that Fe-Co[100] is rotated by $45^\circ$ with respect to [100] direction of MgO, due to the difference in crystalline structure between MgO~(fcc) and Fe-Co~(bcc).

\subsection{Magnetic Characterization}

\begin{figure}[h!]
   \centering
    \includegraphics[width=0.45\textwidth]{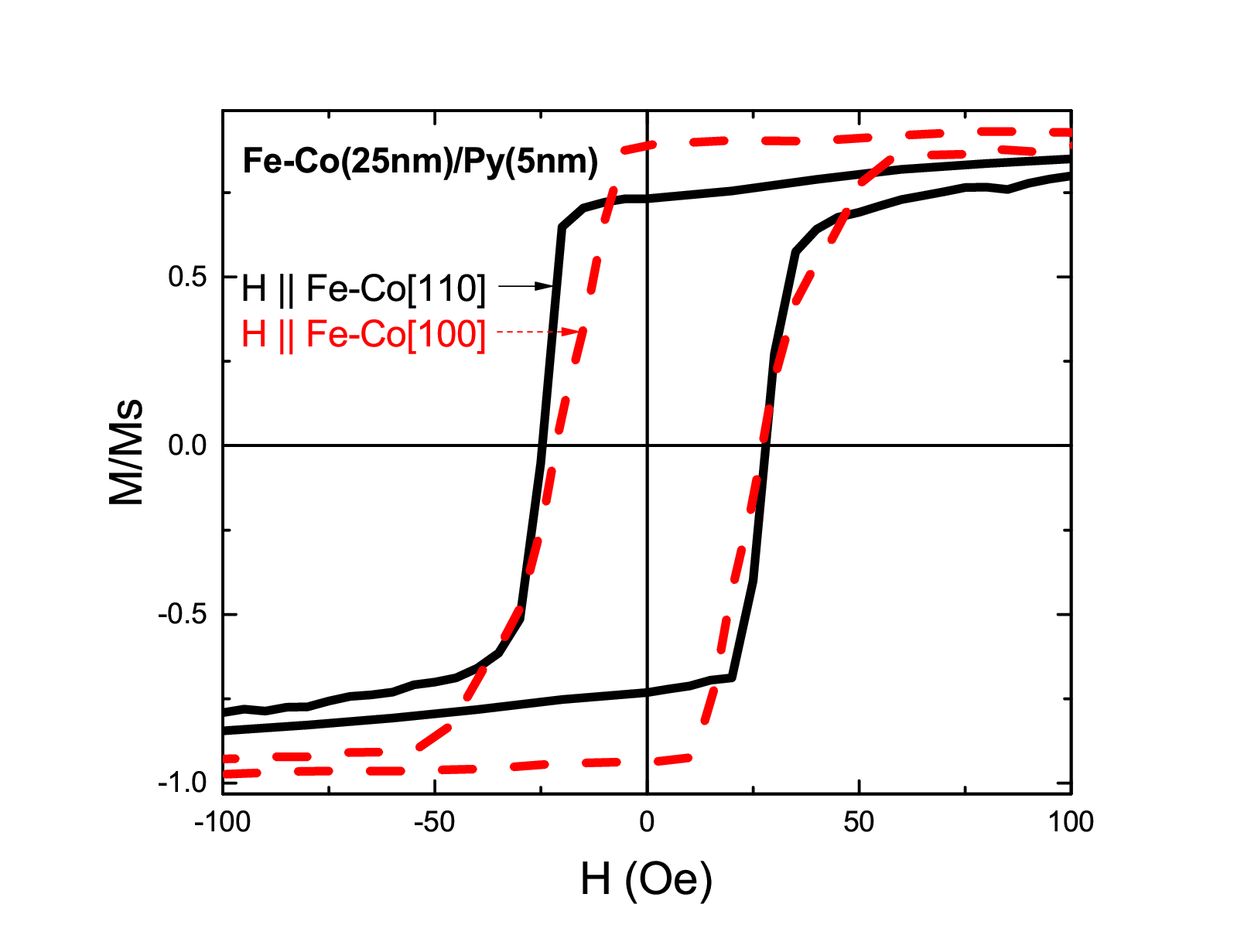}
    \caption{Normalized magnetization vs. applied field for the FeCo25 sample along the hard Fe-Co~[110] and easy Fe-Co~[100] axes.}
    \label{fig:Hysteresys_VSM}
\end{figure}

Magnetic characterization of the samples was performed using vibrating sample (VSM) and magneto-optical Kerr effect (MOKE) magnetometers. The measured in-plane hysteresis loops show different behaviors when the field is applied along the Fe-Co[110] and [100] directions. In Fig. \ref{fig:Hysteresys_VSM}, we show the normalized hysteresis loops measured by VSM for the hard and easy axes on the FeCo25 sample. Along Fe-Co~[110] direction, the hysteresis loops reveal the existence of a hard axis of magnetization, while along the Fe-Co[100] there is an easy axis of magnetization and, as shown in the figure, the remanence magnetization on the hard axis is  smaller. Table \ref{table:coercives} presents the saturation magnetization ($\text{M}_s$) for all samples obtained by VSM and the coercive field, $\text{H}_c$, for the Fe-Co[110] and Fe-Co[100] directions. The $\text{M}_s$ values were calculated by the magnetic moment divided by the total thickness values obtained by XRR.

The hysteresis loops exhibit a continuous profile with no sharp breaks, which suggests a strong coupling between both layers, which a priori have different anisotropies. In all samples we observe the same coercivity along [100] and [110] Fe-Co directions. In the FeCo5 sample we found a lower  coercivity due to the influence of the Py layer; with the exception of the FeCo20 sample that has a lower value. In the case of the two reference samples we obtain a coercive field of 27~Oe for FeCo50/Py3 and 4~Oe for FeCo3/Py50. A coercive permalloy value of $\thicksim 2$~Oe was reported by the authors in Ref. \cite{Py_anisotropy}, for Py thin films, while the value reported for that concentration of Fe-Co is $\thicksim 18$~Oe \cite{velazquez2023bombeo}. Related investigations, such as those carried out in Refs. \cite{cakmaktepe2013underlayer, undelayerFeCoPy} show that Fe-Co grown under a Py underlayer can reduce coercivity due to soft magnetic properties. They suggested that the improvement of soft magnetic properties is caused by a texture change and grain size reduction. In our case, this could be due to the existence of a coupling of Fe-Co with the layer of Py, which causes the film to have a greater or lesser coercive field, depending on the thickness ratio of the film.

From the VSM measurements, we observe that the value of $\text{M}_s$ increases with increasing Fe-Co thickness, as expected if we consider a thickness weighted average. The reported values for pure Fe$_{85}$Co$_{15}$ and Py films are 1700~emu/cm$^3$ \cite{rodriguez2024intrinsic}, and 770~emu/cm$^3$ \cite{JavierGomez}.  

\begin{table}[h!]
\centering
\begin{tabular}{c c c c}
\hline
\hline
 &$M_s(\rm{emu/cm^3})$ & $H^{[110]_{\rm Fe-Co}}_{c}$(Oe) & $H^{[100]_{\rm Fe-Co}}_{c}$(Oe) \\ 
 \hline
 FeCo5 & 913 (10)& 14 (1)& 17 (1) \\ 
 FeCo10 & 1153 (14)& 27 (1)& 28 (1)\\ 
 FeCo15 & 1331 (9)& 26 (1)& 28 (1)\\ 
 FeCo20 & 1453 (7)& 20 (1)& 18 (1)\\ 
 FeCo25 & 1571 (7)& 26 (1)& 24 (1)\\
 FeCo3 & 783 (8)& 4 (1)& 6 (1)\\
FeCo50 & 1686 (7)& 27 (1)& 27 (1)\\
\hline
\hline
\end{tabular}
\caption {Saturation Magnetization and coercive fields for Fe-Co~[110] and [100] directions.}
\label{table:coercives}
\end{table}

\section{Model and Results}
\subsection{Ferromagnetic Resonance Measurements}
\begin{figure}[h!]
    \centering
    \includegraphics[width=0.8\linewidth]{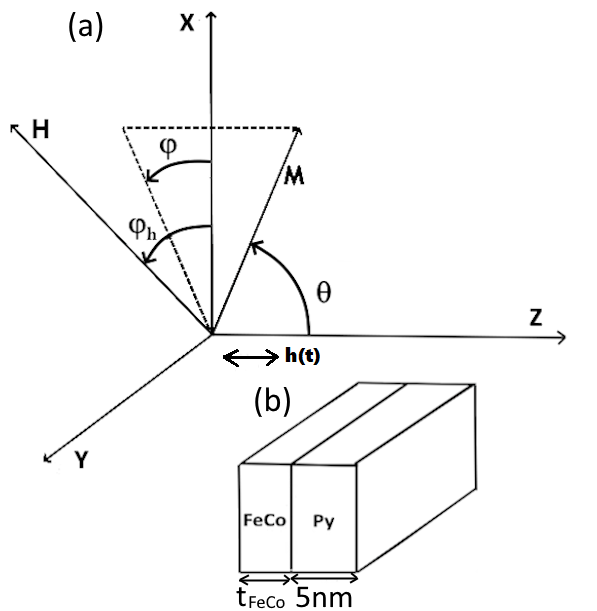}
    \caption{(a) Coordinate system and (b) sketch of the bilayer structure studied in this work with $t_{\rm{FeCo}}$ the thickness of the FeCo layer and 5~nm the thickness of Py.}
    \label{coordenadas}
\end{figure}

In Figs. \ref{coordenadas}(a) and (b), we show the coordinate system used to model the results found in this work and a sketch of the bilayer structure, respectively. The angles $\hat{\theta}$ and $\hat{\varphi}$ are the polar and azimuthal angles of the magnetization, respectively, and $\varphi_{H}$ the angle of magnetic field with respect to the direction of the $x$ axis. The film lies in the $x$-$y$ plane. 

The samples were studied using ferromagnetic resonance (FMR). The in plane angular variation was carried out in X-band, K-band, and Q-band, with a frequency of $\nu$$\thicksim$9.7~GHz, $\nu$$\thicksim$24~GHz and $\nu$$\thicksim$35~GHz, respectively, in a Bruker ESP-300 spectrometer. The samples were placed in the center of a resonant cavity where the derivative of the absorbed power was measured using a standard field modulation and a lock-in detection technique. In all cases, the film plane was perpendicular to the microwave excitation field. Taking into account the observed FMR behavior, we propose the free energy $\text{E}$ per unit area of Eq.~(\ref{free energy}).

\begin{equation}\label{free energy}
\begin{split}
E = \sum_{i=1}^{2} \Big[
& - H M_i t_i \cos(\varphi_i - \varphi_H) \sin\theta_i
+ 2\pi t_i M_i^2 \cos^2\theta_i \\
& - \frac{K_i^c t_i}{4} (\sin^2 2\theta_i + \sin^2 2\varphi_i \sin^4\theta_i) \\
& - K_i^u t_i \sin^2\theta_i \cos^2(\varphi_i - \pi/4)
\Big] \\
& - J_{\text{ex}} [\cos\theta_1\cos\theta_2 + \sin\theta_1\sin\theta_2\cos(\varphi_1 - \varphi_2)].
\end{split}
\end{equation}

Here, $i$ runs over the two magnetic layers in the system. The first term is the Zeeman energy, where $\text{H}$ is the applied external field, $\text{M}_i$ and $\text{t}_i$ the saturation magnetization and the thickness of the $i$-th layer, respectively. The second one is related to the sample shape, the third and fourth terms are the cubic magnetocrystalline anisotropy  of magnitude  $\text{K}^c$ and uniaxial anisotropy $\text{K}^u$ contributions to free energy, respectively. Finally, the last term corresponds to the exchange energy, with $J_{\text{ex}}$ being the bilinear exchange constant. The sign of $J_{\text{ex}}$ is chosen so that it is positive (negative) for a parallel (antiparallel) coupled system.

Eq.~\eqref{Landau-Lifshitz-Gilbert} is the Landau-Lifshitz-Gilbert equation of motion with Gilbert damping \cite{Gilbert}, where {\bf M}$_i$ stands for the magnetization of each layer in spherical coordinates, being ${\bf{M}}_i=\text{M}_{0_{i}}\hat{e}_r+ m_{\theta_i}\hat{e}_{\theta_i}+m_{\varphi_i}\hat{e}_{\varphi_i}$, where $m_\theta$ and $m_\varphi$ are the components of the magnetization and $\text{M}_{0_{i}}$ the saturation magnetization of each layer. For the equilibrium orientation and retaining only terms to the first order of $m_\theta$ and $m_\varphi$, we obtain the set of four equations condensed in Eq.~\eqref{bilayer}. The resonance field ($\text{H}_r$) values can be extracted as a function of the in plane angle of the external applied field using the Smit and Beljers formalism for a bilayer system \cite{zhang1994angular, layadi1990ferromagnetic}.

\begin{equation}
\label{Landau-Lifshitz-Gilbert}
\frac{\partial {\bf{M}_i}}{\partial {t}}=-\gamma \bf{M}_i \times \bf{H} - \frac{\gamma\alpha_i}{M_i}[\bf{M}_i(\bf{M}_i \cdot \bf{H}) -\bf{H}M_i^2]
\end{equation}

\begin{equation}
\label{bilayer}
\begin{pmatrix}
A+B
\end{pmatrix}
\begin{pmatrix}
m_{\theta_1}\\
m_{\varphi_1} \\
m_{\theta_2} \\
m_{\varphi_2} 
\end{pmatrix}
=
\begin{pmatrix}
0\\
-M_1h_0e^{i\omega_0t} \\
0 \\
-M_2h_0e^{i\omega_0t} 
\end{pmatrix},
\end{equation}

where:

\begin{widetext}
\begin{equation*}
{\Large
A=
\begin{pmatrix}
i\frac{\omega}{\gamma}+\frac{E_{\theta_1\varphi_1}}{t_1M_1\sin\theta_1}&\frac{E_{\varphi_1\varphi_1}}{t_1M_1\sin^2\theta_1}&\frac{E_{\theta_2\varphi_1}}{t_1M_2\sin\theta_1}& \frac{E_{\varphi_1\varphi_2}}{t_1M_2\sin\theta_1\sin\theta_2}\\
\frac{-E_{\theta_1\theta_1}}{t_1M_1}& i\frac{\omega}{\gamma}-\frac{E_{\theta_1\varphi_1}}{t_1M_1\sin\theta_1}& \frac{-E_{\theta_1\theta_2}}{t_1M_2}& \frac{-E_{\theta_1\varphi_2}}{t_1M_2\sin\theta_2}\\
\frac{E_{\theta_1\varphi_2}}{t_2M_1\sin\theta_2}& \frac{E_{\varphi_1\varphi_2}}{t_2M_1\sin\theta_1\sin\theta_2}& i\frac{\omega}{\gamma}+\frac{E_{\theta_2\varphi_2}}{t_2M_2\sin\theta_2}& \frac{E_{\varphi_2\varphi_2}}{t_2M_2\sin^2\theta_2}\\
\frac{-E_{\theta_1\theta_2}}{t_2M_1}& \frac{-E_{\theta_2\varphi_1}}{t_2M_1\sin\theta_1}& \frac{-E_{\theta_2\theta_2}}{t_2M_2}& i\frac{\omega}{\gamma}-\frac{E_{\theta_2\varphi_2}}{t_2M_2\sin\theta_2}

\end{pmatrix},
}
\end{equation*}
\begin{equation*}
{\Large
B=
\begin{pmatrix}
\frac{\alpha_1E_{\theta_1\theta_1}}{t_1M_1}&\frac{\alpha_1E_{\theta_1\varphi_1}}{t_1M_1\sin\theta_1}&\frac{\alpha_1E_{\theta_1\theta_2}}{t_1M_2} & \frac{\alpha_1E_{\theta_1\varphi_2}}{t_1M_2\sin\theta_2}\\
\frac{\alpha_1E_{\theta_1\varphi_1}}{t_1M_1\sin\theta_1}\! & \frac{\alpha_1E_{\varphi_1\varphi_1}}{t_1M_1\sin^2\theta_1}\! & \frac{\alpha_1E_{\varphi_1\theta_2}}{t_1M_2\sin\theta_1\!} & \frac{\alpha_1E_{\varphi_1\varphi_2}}{t_1M_2\sin\theta_1\sin\theta_2}\!\\
\frac{\alpha_2E_{\theta_1\theta_2}}{t_2M_1}\! & \frac{\alpha_2 E_{\theta_2\varphi_1}}{t_2M_1\sin\theta_1}\! & \frac{\alpha_2E_{\theta_2\theta_2}}{t_2M_2}\! & \frac{\alpha_2E_{\theta_2\varphi_2}}{t_2M_2\sin\theta_2}\!\\
\frac{\alpha_2E_{\theta_1\varphi_2}}{t_2M_1\sin\theta_2}\! & \frac{\alpha_2E_{\varphi_1\varphi_2}}{t_2M_1\sin\theta_1\sin\theta_2}\! & \frac{\alpha_2E_{\theta_2\varphi_2}}{t_2M_2\sin\theta_2}\! &\frac{\alpha_2E_{\varphi_2\varphi_2}}{t_2M_2\sin^2\theta_2}\!
\end{pmatrix}.
}
\end{equation*}
\end{widetext}

In Eq.~\eqref{Landau-Lifshitz-Gilbert}, $\bf{H}$ is expressed as $\bf{H}_{ext}+\bf{h}(t)$, with $\bf{H}_{ext}$ the external field and $\bf{h}(t)$ the microwave field, expressed as ${\bf{h_0}} e^{i\omega_0t}$; and it is applied along the z direction, perpendicular to $\bf{M}$ and $\bf{H}$, the ${\bf{h_0}}$ value was set to 1~Oe for the experiment. $\gamma$ is the giromagnetic ratio, because the g-factor is almost the same for Fe-Co and Py, we assume $18.36 \times 10^{6}{(\text{Oe} \cdot \text{s})}^{-1}$ \cite{rodriguez2024intrinsic} for both layer, $\omega$ is the excitation angular frequency, and $\alpha_i$ the damping parameter. For the latter, values of 0.0026 \cite{rodriguez2024intrinsic} for the Fe-Co layer and 0.0083 for the Py layer were used \cite{JavierGomez}. The subscripts in E indicate second derivatives with respect to the angular coordinates of the 
i-th layer: $E_{\theta_i\varphi_i}=\frac{\partial^2E}{\partial\theta_i\partial\varphi_i}$, $E_{\theta_i\theta_i}=\frac{\partial^2E}{\partial\theta_i\partial\theta_i}$, $E_{\varphi_i\varphi_i}=\frac{\partial^2E}{\partial\varphi_i\partial\varphi_i}$. The equilibrium angles are obtained by minimizing the free energy with respect to the angles $\theta_i$ and $\varphi_i$. In this case $\theta_i=\frac{\pi}{2}$, which means that $\text{M}_i$ lies in the plane of the film. The values used for the $\text{M}_s$ of each layer were 1700~emu/cm$^3$ for Fe-Co and 770~emu/cm$^3$ for Py, as mentioned in the previous section. 

\subsection{Model}
\subsubsection{Dispersion relations and eigenvectors}

\begin{figure}[h!]
   \centering
    \hspace{-0.8cm}\includegraphics[width=0.45\textwidth]{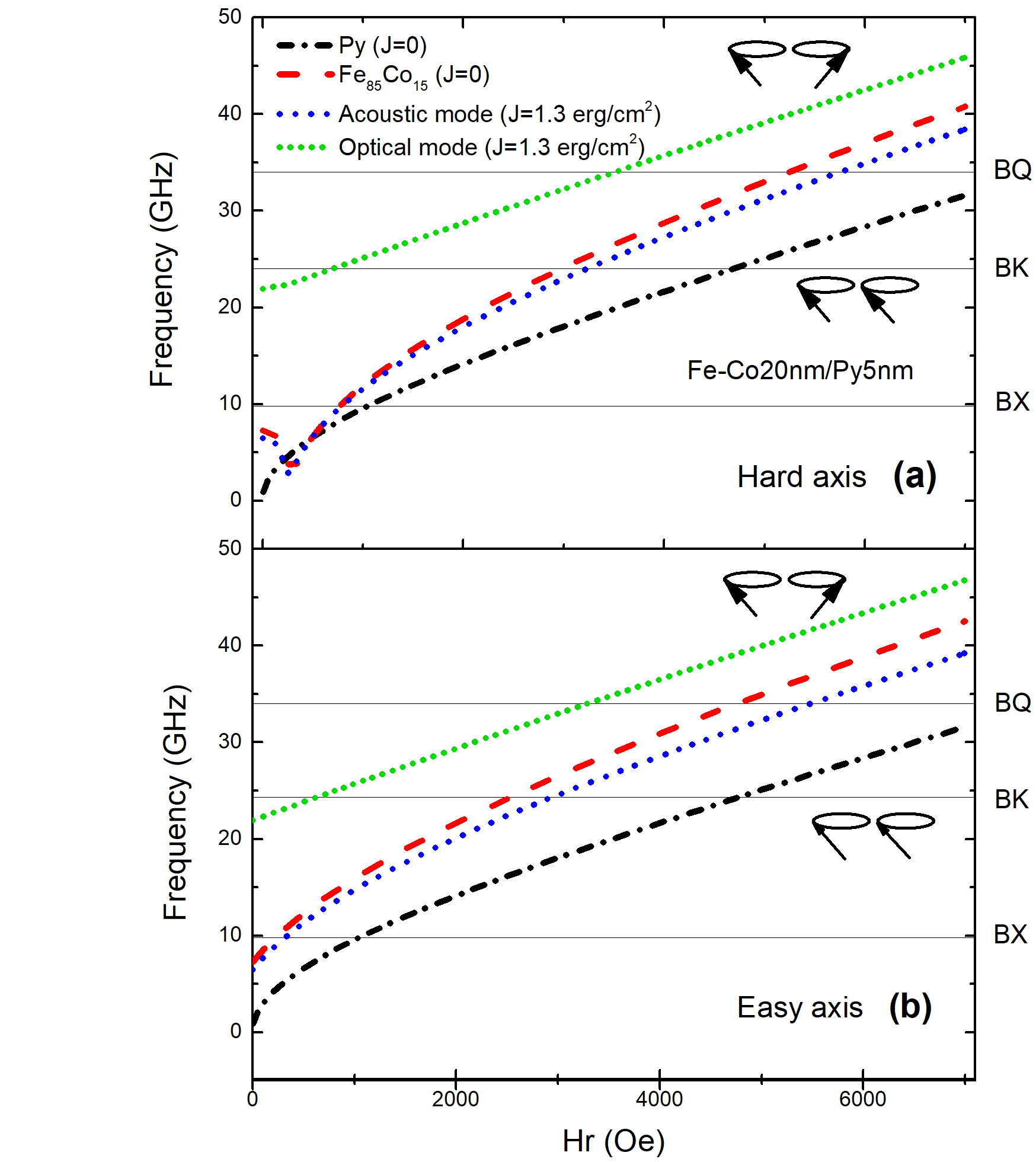}
    \caption{Dispersion relations of coupled and uncoupled systems along the (a) hard axis and (b) easy axis for the FeCo20 sample. Horizontal lines correspond to different excitation frequencies used in this work.}
    \label{relacdisp}
\end{figure}

Using the coupled bilayer model, the dispersion relations were simulated, finding the roots of the determinant of the matrix $(A + B)$ in Eq.~(\ref{bilayer}). The determinant has four solutions, two are negative values of $\omega$, and do not correspond to any physical resonance modes. In Fig. \ref{relacdisp}, the dispersion relations for the FeCo20 sample are plotted for the coupled and uncoupled system using the exchange constant that fits the model, along the hard and easy axes of magnetization. The horizontal lines are the frequencies used in the FMR experiments. 

If the system were uncoupled, two independent modes would be seen, one corresponding to the precession of Py at a higher field for a given frequency, and the other, the Fe-Co mode, at lower fields. When the system is ferromagnetically coupled, there are also two modes, the acoustic mode, at higher fields, corresponding to the rf components of magnetization resonating in phase, and the optical mode at lower fields, that corresponds to the rf components of magnetization resonating out of phase. In Fig.~\ref{relacdisp} we can observe the frequency gap of $\thicksim15$~GHz that prevents the observation of this mode at low frequencies. However, this mode exhibits low intensity and could not be detected in the experimental measurements at high frequencies.

\subsubsection{Anisotropy constants}

\begin{figure}[h!]
   \centering
   \hspace{-0.8cm}
    \includegraphics[width=0.35\textwidth]{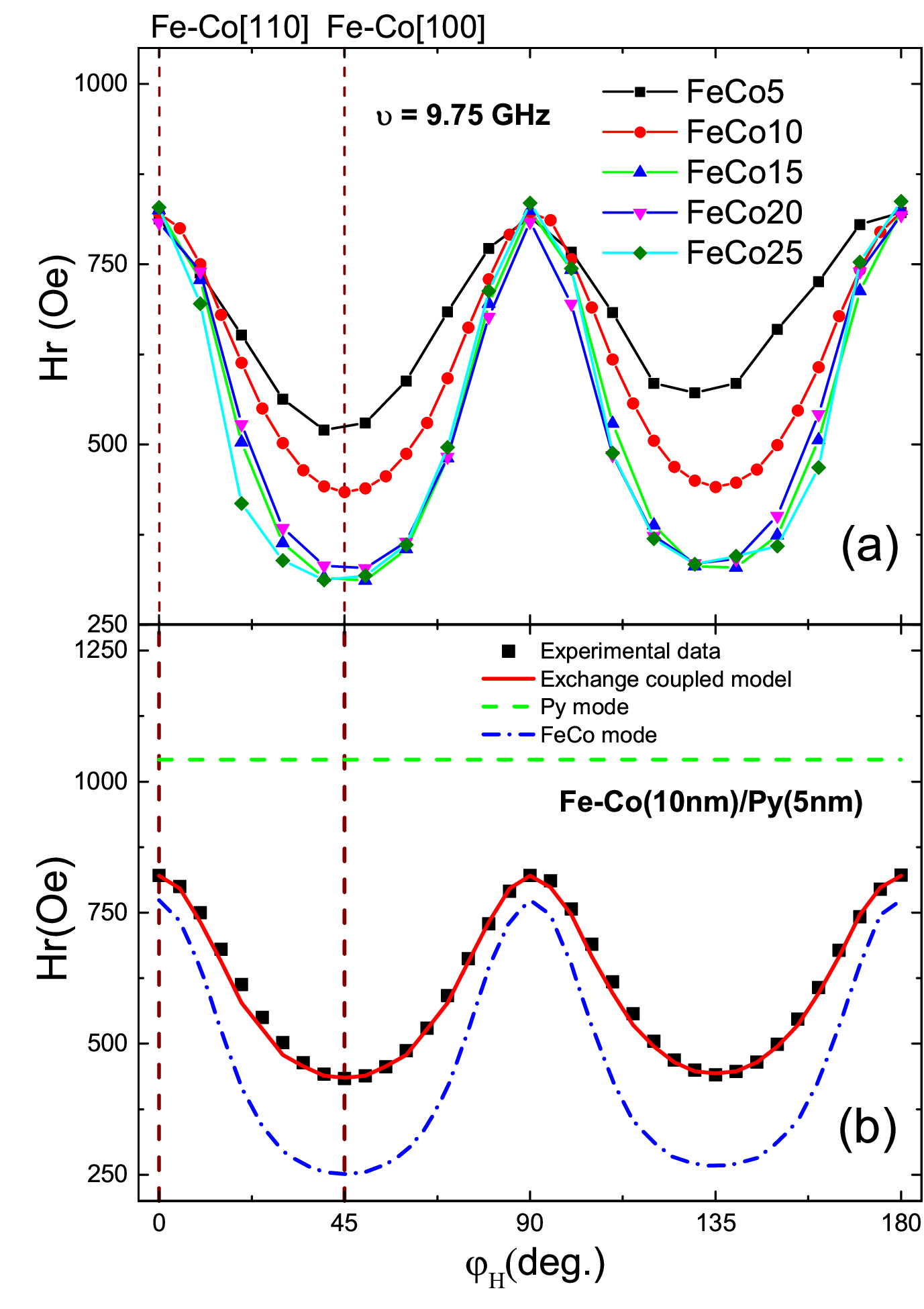}
    \caption{(a) $\text{H}_r$ vs $\varphi_H$ for all the samples in the experiment, (b) $\text{H}_r$ vs $\varphi_H$ for a coupled bilayer, and for Py and Fe-Co layers (no-coupling) .}
    \label{ang_var}
\end{figure}

The FMR in-plane angular variation was carried out between $\varphi_H=0^\circ$ and $\varphi_H=180^\circ$, and experimental data were fitted using the exchange coupled model given in Eq.~(\ref{bilayer}), using the thicknesses and values of $\text{M}_s$ for each layer presented in Table~\ref{table:coercives}. The values of $\varphi_i$ were numerically calculated by minimizing the free energy equation for each value of $\text{H}$ and $\varphi_H$. In Fig.\ref{ang_var}(a) the in plane angular variation is plotted for all the samples, the resonance fields along the hard axis are approximately the same for all samples, whereas along the easy axis the resonance fields differ and depend on the differences in anisotropy, also observing that when the thickness of Fe-Co increases, the $\text{H}_r$ vs. $\varphi_H$ curve stretch more, due to an increase in the cubic anisotropy constant of the three thicker films compared to the two thinner ones. From the best fit with the experimental data we extract the anisotropy constants expressed in magnetic field units $\text{H}^u$, $\text{H}^c$ which follow the relation $\text{H}^{u,c}=\frac{2\text{K}^{u,c}}{\text{M}}$, as well as $J_\text{ex}$. $\varphi_H=0^\circ$ corresponds to $\bf{H}_{ext}$~||~Fe-Co~[110] direction. In Fig.~\ref{ang_var}, the in plane angular variation shows maxima at $\varphi_H=0^\circ$ (Fe-Co~[110]), $\varphi_H=90^\circ$ and $\varphi_H=180^\circ$, while the minima are at $\varphi_H=45^\circ$ ([Fe-Co~100]) and $\varphi_H=135^\circ$. This four-fold symmetry is characteristic of the angular variation around a [001] axis in system with cubic magnetocrystaline anisotropy. The Fe-Co [100] direction is rotated by $45^\circ$ with respect to [100] direction of MgO, and for this concentration of Fe-Co it is coincident with the cubic easy axis of magnetization \cite{rodriguez2024intrinsic, Muhge}. The measurerements of FMR and VSM suggest the idea of epitaxial growth on the MgO substrate, due to the four-fold behavior observed.

As we can see in Fig.~\ref{ang_var}, the minimum values, located at $\varphi_H=45^\circ$ are not equal, which indicates the presence of a uniaxial anisotropy, which is much smaller than the cubic anisotropy and is possibly induced by the fabrication method. The uniaxial easy axis at $\varphi=45^\circ$ matches with the cubic easy axis of magnetization, which coincides with the results obtained by VSM and MOKE.

Ref.~\cite{yin2006magnetocrystalline} studied Py (at the concentration studied in this work) and obtained that the cubic anisotropy was extremely small, due to the Fe-Ni bonding that quenches the magnitude of $\text{K}^c$. In our model we then assume that Py anisotropy is $\text{K}^c$=0 and the magnetocrystalline contribution to the anisotropy in the bilayers is only due to the Fe-Co layer. The resonance field for all samples in the experiment is between the values for Py and for Fe-Co single layers. In Fig. \ref{ang_var}(b) the calculated in-plane angular variation of $\text{H}_r$ is plotted with green and blue lines, for the single layer of Py and Fe-Co respectively. The fact that only one resonance peak is seen in X band means that the system is coupled, and only the in phase (acoustic mode) can be seen at this frequency. The linewidth ($\Delta \text{H}$) along the hard axis direction has values about 23~Oe, similar than the reported by the authors in Ref.\cite{rodriguez2024intrinsic}, except for samples of 5~nm and 10~nm, with values 41~Oe and 50~Oe respectively. These relative low values for the three thicker films may lead to the idea of better quality samples with respect to the thinner ones. In Table \ref{table:anisotropias}, the values of anisotropy and exchange constants extracted from the model for each sample are shown.

The values for $\text{H}^c$ increase with the Fe-Co thickness, going from 216~Oe for the thinnest film, reaching values in the range (283-308) Oe for larger thicknesses of Fe-Co. This behavior could be attributed to a thickness-weighted contribution of the anisotropy field from each layer, which leads to a reduction in the effective $\text{H}_c$ as Fe-Co thickness decreases. Moreover, this behavior could be associated to the thickness dependence of the magnetocrystalline anisotropy field, which decreases as the thickness decreases \cite{pelzl2003spin}, our values are comparable to those obtained by \cite{rodriguez2024intrinsic, saba2025magnetization}. 

The exchange constant in these bilayers is a parameter that depends on the interface between the two ferromagnets \cite{alsmadi2014interfacial}. In Table~\ref{table:anisotropias}, we show the values of $J_{{\text{ex}}}$ obtained from the fittings. We have found similar values in all samples, $\langle J_{\text{ex}} \rangle\thicksim$~1~erg/cm$^2$, and $J_{{\text{ex}}}$ which is expected, since all samples were fabricated following the same protocol, and thus a similar interface quality is anticipated.

\begin{table}[h!]
\centering
\begin{tabular}{c c c c }
\hline
\hline
  & $\text{H}^u$(Oe) & $\text{H}^c$(Oe) & $J_{ex}(\rm{erg/cm^2})$\\ 
 \hline
  FeCo5  & 42 & 216 & >1.2\\ 
  FeCo10 & 6 & 262 & >0.8\\ 
  FeCo15 & 10 & 308 & >1.2\\ 
  FeCo20 & 7 & 283 & >1.3\\ 
  FeCo25 & 14 & 298 & >0.8\\ 
 \hline
 \hline
\end{tabular}
\caption {Values of anisotropy and exchange constants calculated using the bilayer model.}
\label{table:anisotropias}
\end{table}

\begin{figure}[h!]
    \includegraphics[width=0.45\textwidth]{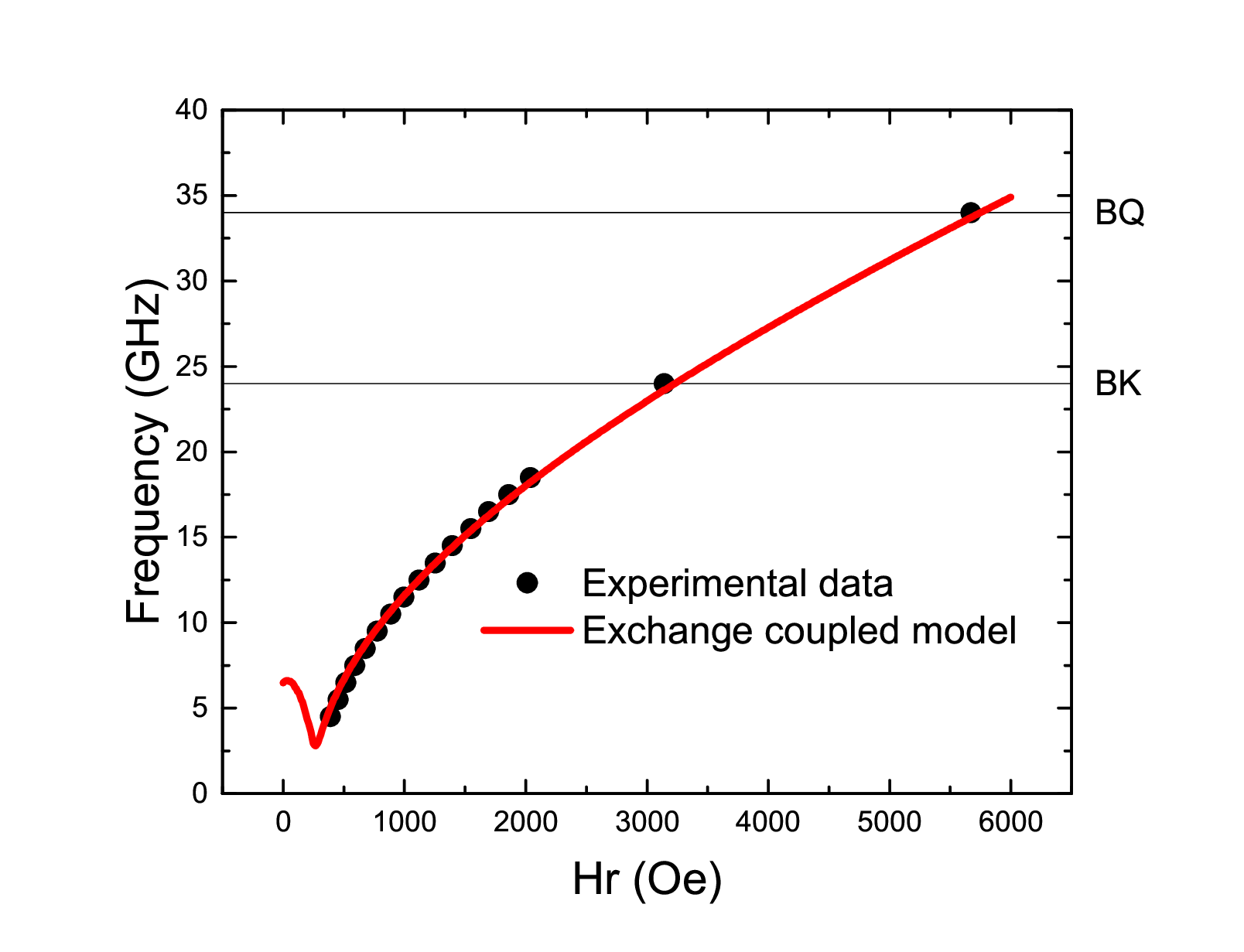}
    \caption{Dispersion relation from broadband experiments for the FeCo20 sample.}
    \label{VNA}
\end{figure}

We have also carried out experiments of broad band FMR. The resonance field values were measured only along the hard axis of magnetization of the samples for frequencies from 3.5~GHz to 19.5~GHz using a Vector Network Analyzer (VNA) and a coplanar waveguide, in which the transmitted signal $S_{\text{21}}$ is detected. Fig.~\ref{VNA} shows the experimental results for the FeCo20 sample, fitted using the bilayer exchange model, for the case of the acoustic mode observed in resonance. In the figure we can corroborate that our model fits with very good accuracy with the resonance fields obtained in the broad band FMR experiment, using the same parameters that we use to fit the in-plane FMR angular variation.

\subsection{Components of magnetization and spin currents}

\begin{figure}[h!]
\begin{subfigure}[h]{0.45\textwidth}
    \includegraphics[width=0.9\linewidth, height=6cm]{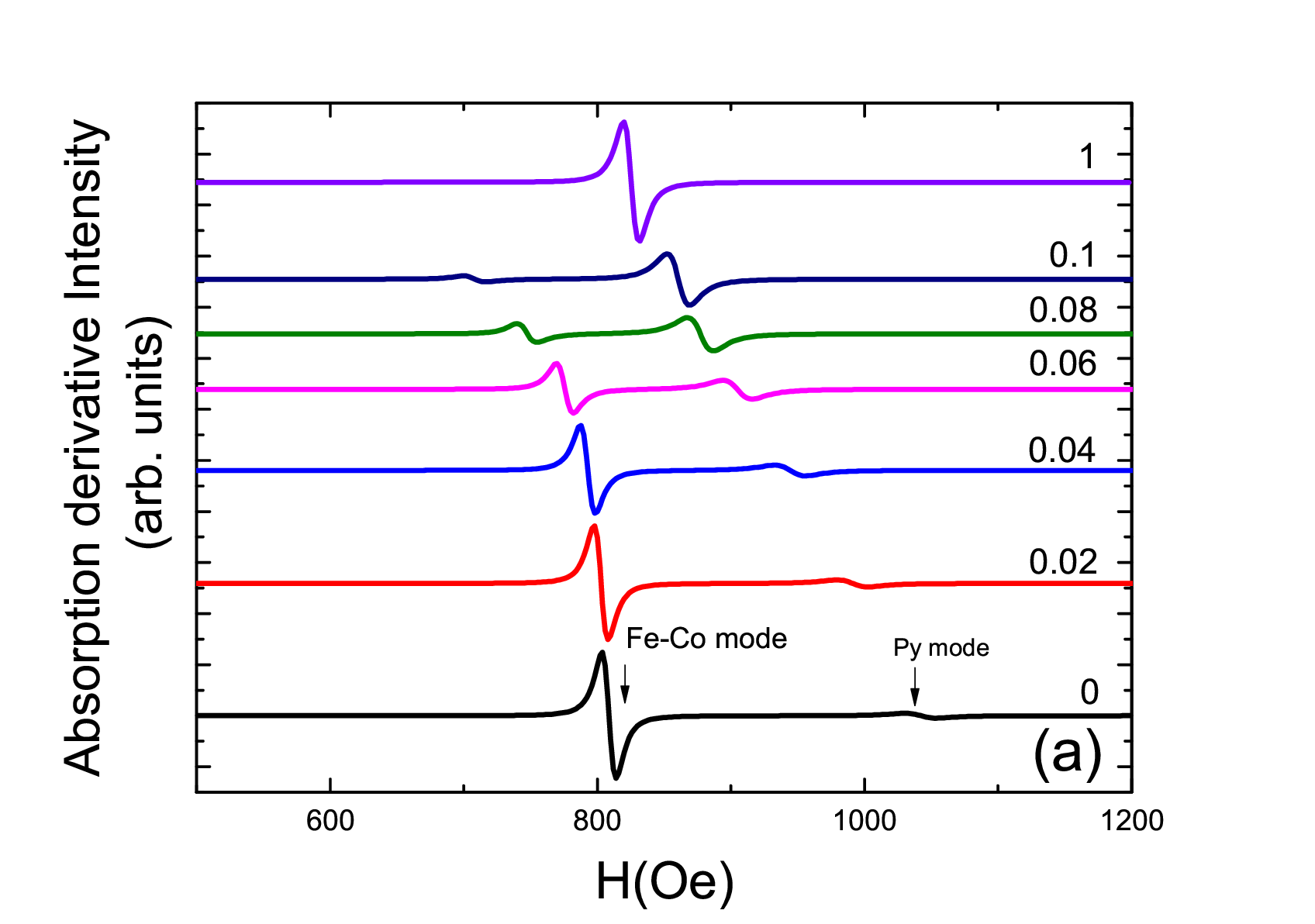} 
\end{subfigure}
\begin{subfigure}[h]{0.48\textwidth}
    \includegraphics[width=0.9\linewidth, height=6cm]{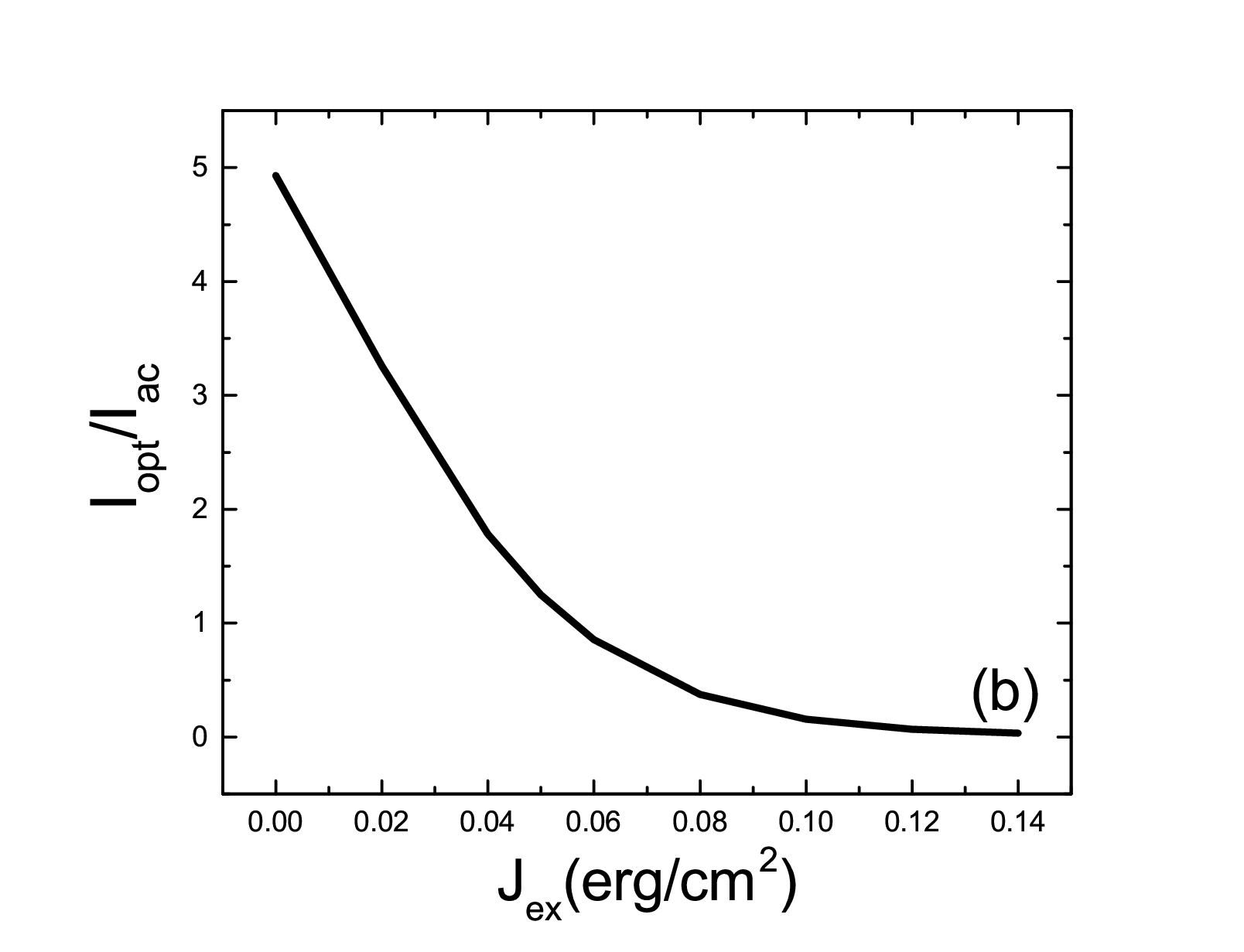}
\end{subfigure}

\caption{(a) Simulated ferromagnetic resonance spectra measured along the hard axis direction for different values of $J_{{\text{ex}}}$ in the FeCo25 sample, (b) Intensity ratio (I$_{\rm opt}$/I$_{\rm ac}$) as a function of $J_{{\text{ex}}}$. Both panels correspond to calculations performed at X-band frequency.}
\label{Intensity}
\end{figure}

If we solve the inhomogeneous 4x4 system of Eq.~(\ref{bilayer}), for each excitation frequency, the results are the components of rf-magnetization which, expressed in matrix form, is a vector with components
$\begin{pmatrix}
m_{\theta_1},m_{\varphi_1},m_{\theta_2},m_{\varphi_2} 
\end{pmatrix}$. If the system is uncoupled, the solutions will be 
$\begin{pmatrix}
m_{\theta_1},m_{\varphi_1},0,0 
\end{pmatrix}$ 
for Fe-Co layer, and
$\begin{pmatrix}
0,0,m_{\theta_2},m_{\varphi_2} 
\end{pmatrix}$ 
for Py layer. When a $J_{\text{ex}}$ is present, magnetization components of one layer start to mix with the other one, obtaining the acoustic and optical coupled modes. As $J_{{\text{ex}}}$ increases, both the optical and acoustic modes shift toward lower resonance fields, while the mode intensity ratio (I$_{\rm opt}$/I$_{\rm ac}$) decreases. In Fig.~\ref{Intensity}, resonance field spectra and intensity ratios calculated for different values of $J_{{\text{ex}}}$ are simulated for the FeCo25 sample. As shown in Fig.~\ref{Intensity}~(a), the optical mode progressively vanishes as $J_{{\text{ex}}}$ increases, whereas the acoustic mode approaches the resonance field value corresponding to $J_{{\text{ex}}}\rightarrow\infty$. The fact that the $\text{H}_\text{r}$ measured experimentally along the hard axis (corresponding to $J_{{\text{ex}}}\approx1~\text{erg/cm}^2$) is close to the value obtained for $J_{{\text{ex}}}\rightarrow\infty$ suggest an intrinsic uncertainty in the determination of $J_{{\text{ex}}}$. This interpretation is further supported by the results shown in Fig.~\ref{ang_var}~(a) where the measured $\text{H}_\text{r}$ values along the hard axes are nearly identical. For these reasons, $J_{{\text{ex}}}$ can only be determined in terms of a lower bound, as reported in the Table~\ref{table:anisotropias}. Fig.~\ref{Intensity}~(b) illustrates the ratio between the optical and acoustic modes as a function of $J_{\text{ex}}$, extracted from the peak intensities in Fig.~\ref{Intensity}~(a).

\begin{figure}[h!]
\begin{overpic}[width=0.55\textwidth]{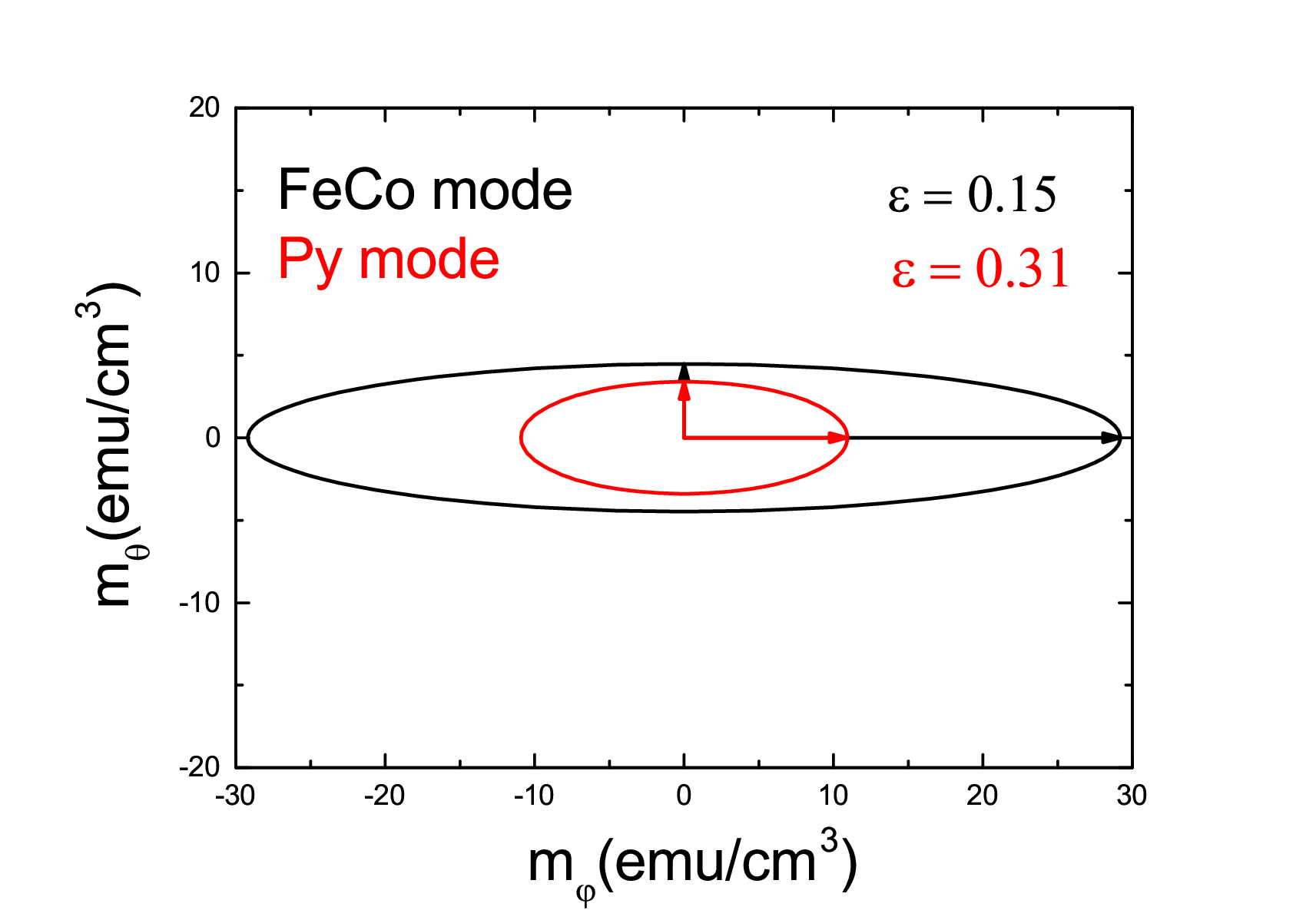}
\put(19,20){%
\begin{minipage}{0.75\linewidth}
\includegraphics[width=\linewidth]{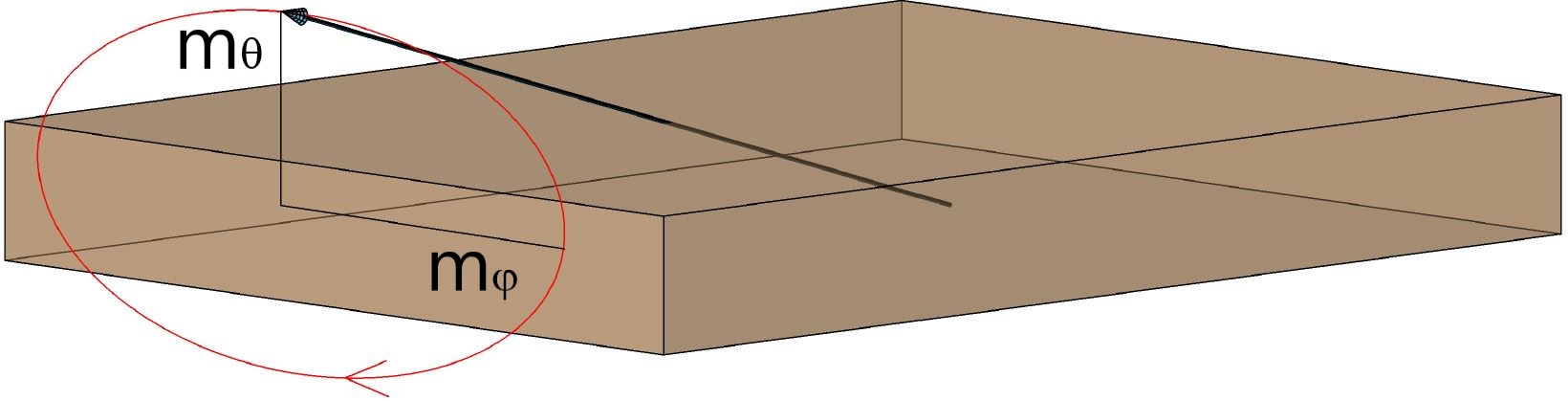}
\end{minipage}
}
\end{overpic}
\caption{Magnetization precession trajectory for uncoupled layers of Fe-Co and Py modes. inset: Sketch of the magnetization precession of each layer in the film.}
\label{ellipticity}
\end{figure}

A parameter that defines the magnetization-precession trajectory is the ellipticity $\varepsilon$ which is defined as $\varepsilon = {m_\theta}/{m_\varphi}$. In Fig.~\ref{ellipticity}, we show the ellipsoid generated by the magnetization precession for Py and Fe-Co in the uncoupled case, being $\varepsilon_{\rm Fe-Co} = 0.15$ and $\varepsilon_{\rm Py} = 0.31$. The ellipticity computed with the model is in agreement with the results obtained in Ref.~\cite{ando2011inverse} for the ellipticity and is an indication that the proposed model is feasible for the estimation of the magnetization components.

\begin{figure}[h!]
\begin{subfigure}{0.45\textwidth}
    \includegraphics[width=0.9\linewidth, height=6cm]{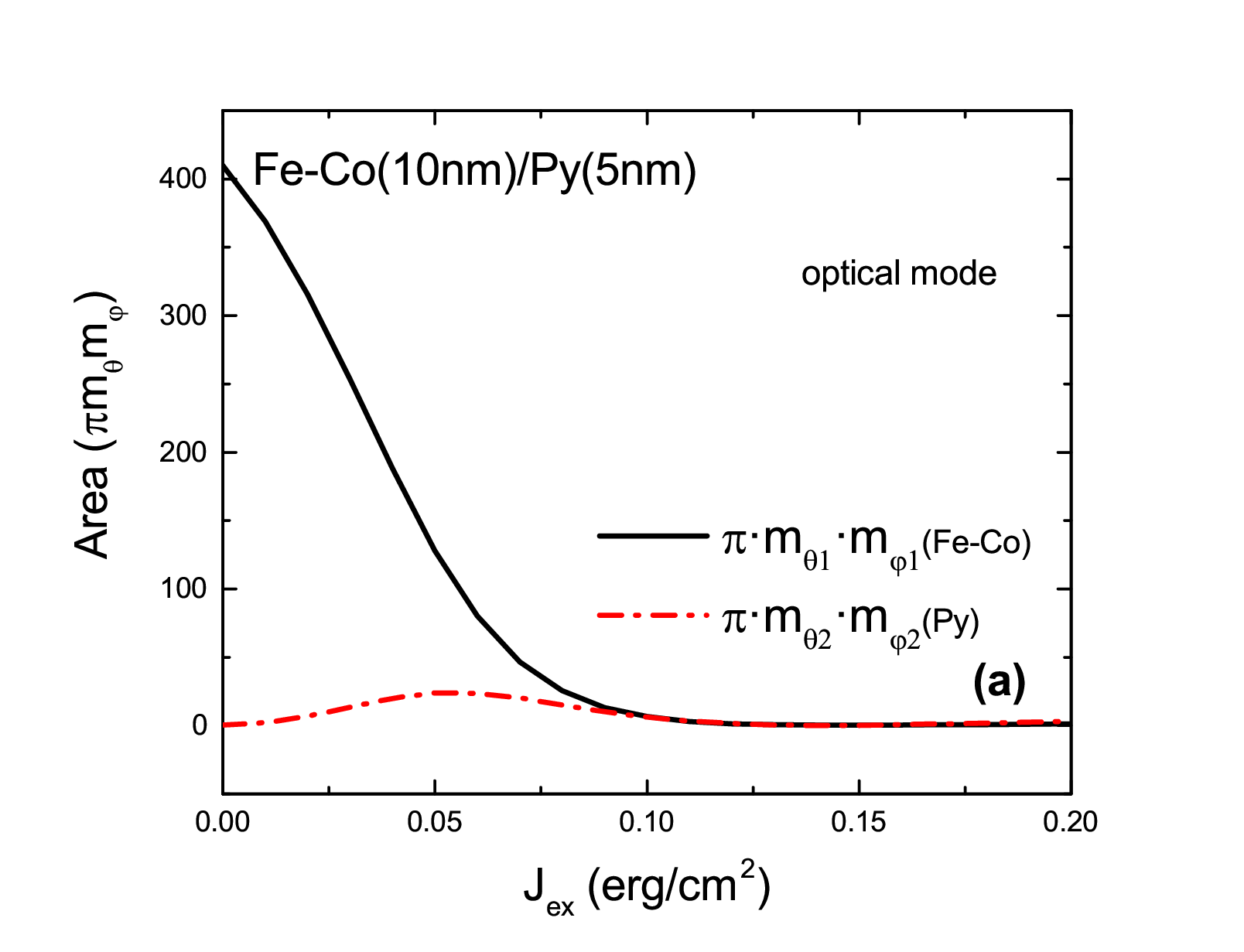} 
\end{subfigure}
\hfill
\begin{subfigure}{0.45\textwidth}
    \includegraphics[width=0.9\linewidth, height=6cm]{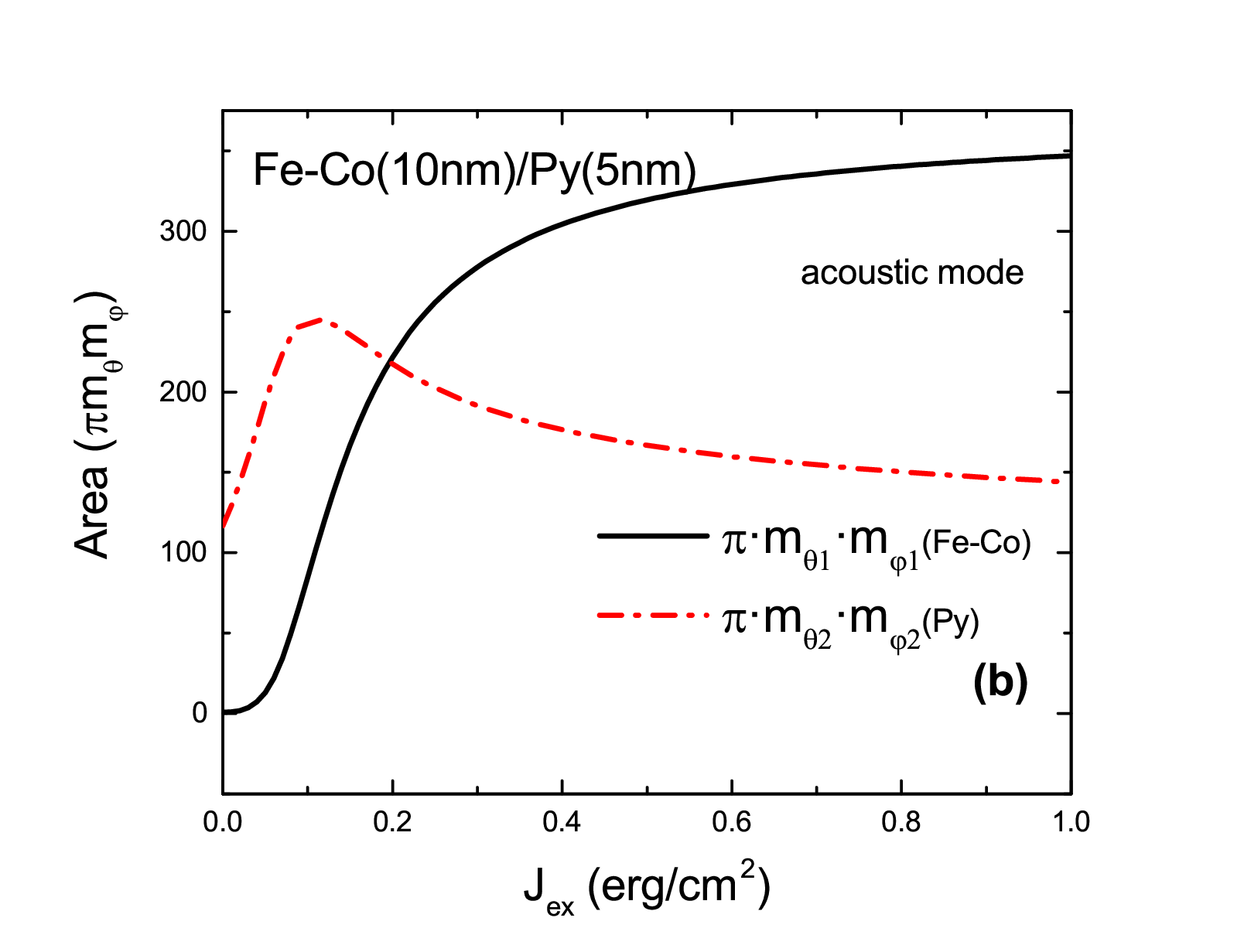}
\end{subfigure}
\caption{Area of magnetization precession trajectory for the Fe-Co and Py components along the hard axis, for the (a) optic, and (b) acoustic modes.}
\label{area}
\end{figure}

\begin{figure}[h!]
\begin{overpic}[width=0.5\textwidth]{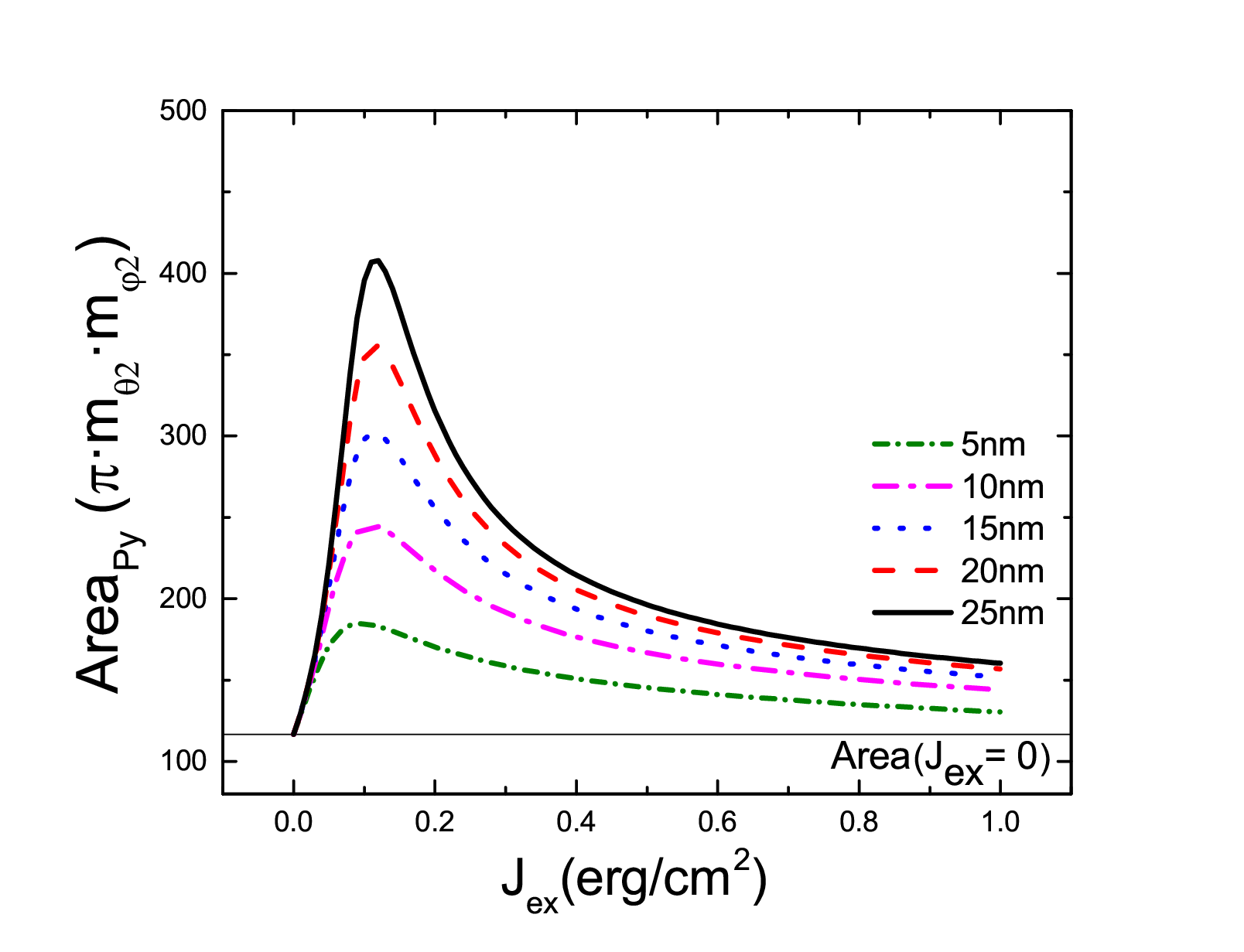}
\put(40,53.8){%
\begin{minipage}{0.46\linewidth}
\includegraphics[width=\linewidth]{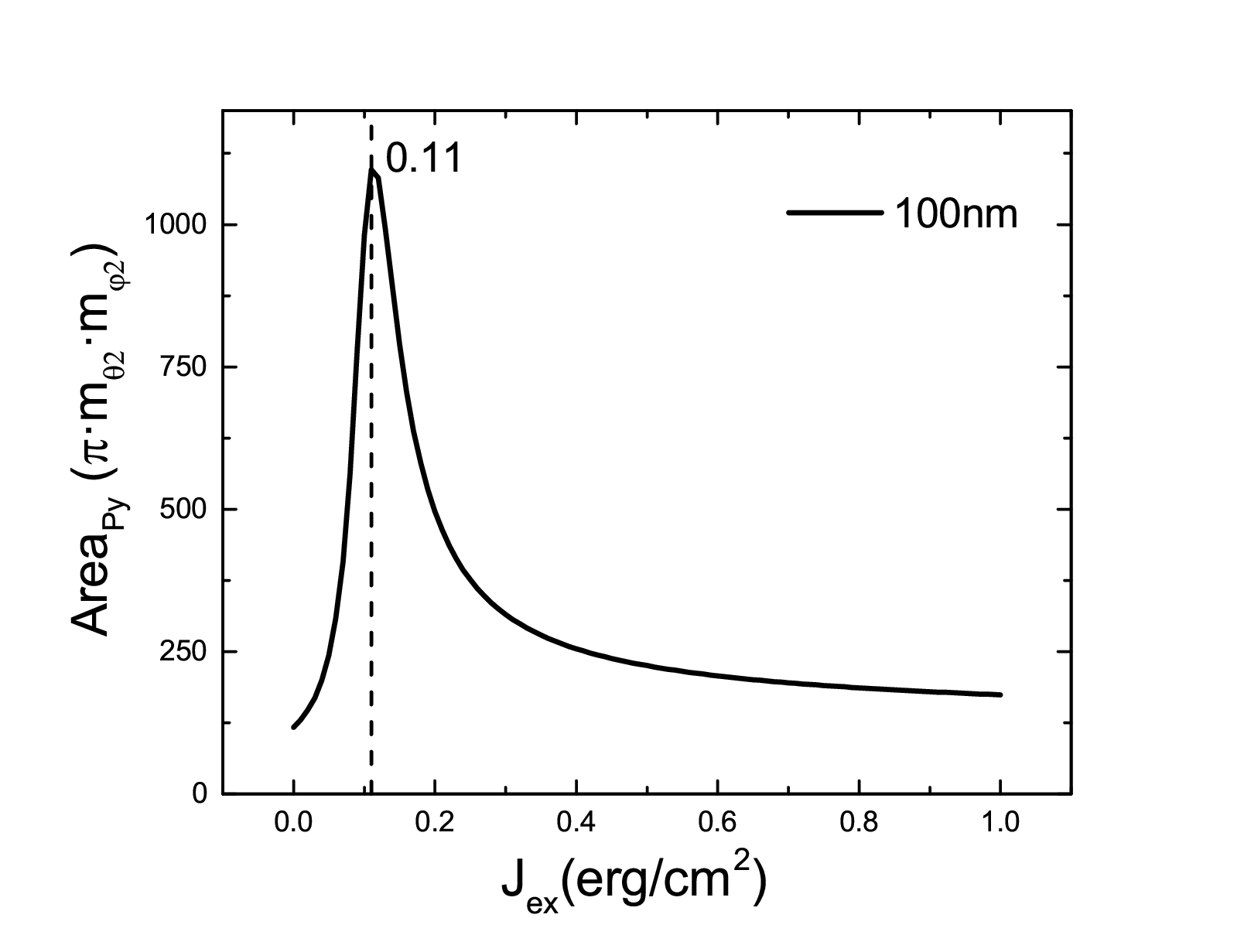}
\end{minipage}
}

\end{overpic}
\caption{Contribution of the Py layer for different samples with different thicknesses of the FeCo layer (5\,nm, 10\,nm, 15\,nm, 20\,nm and 25\,nm), and 5\,nm of Py, inset: 100\,nm FeCo sample. All simulations were made along the hard axis.}
\label{contrPy_all}
\end{figure}

In Ref.~\cite{ando2011inverse}, the authors demonstrated that a parameter that defines the magnetization precession trajectory is given by the elliptical area, defined as $\text{S}=\pi m_\theta m_\varphi$. They found that the spin current amplitude injected in an adjacent metallic layer is maximized when this area is maximized. For this bilayer system, with magnetization components
$\begin{pmatrix}
m_{\theta_1},
m_{\varphi_1},
m_{\theta_2},
m_{\varphi_2} 
\end{pmatrix}$, 
the area of the magnetization precession trajectory can be separated in two components, the precession area of the Fe-Co layer
$\text{S}_\text{{Fe-Co}}~=~\pi m_{\theta1} m_{\varphi1}$, and that of the Py layer $\text{S}_\text{{Py}}=\pi m_{\theta2} m_{\varphi2}$. In our simulations, the magnetization components were calculated for all the samples along the easy and hard axes of magnetization, and hence the contribution to the elliptical area of magnetization for both layers could be estimated. For instance, for FeCo10 along the hard axis of magnetization, for the acoustic mode, as the exchange constant increases, the area of Fe-Co layer also increases from $\text{S}_\text{{Fe-Co}}$~=~0, for $J_{\text{ex}}=0$, to a value corresponding to $\text{S}_\text{{Fe-Co}}\underset{J_{{\text{ex}}}\rightarrow \infty}{\longrightarrow}356$~({emu/cm$^3$})$^2$. The Py area starts with the value of Py single layer precession area, $\text{S}_\text{{Py}}$~=~116 ({emu/cm$^3$})$^2$ corresponding to $J_{\text{ex}}=0$. As $J_{\text{ex}}$ increases, the contribution of Py increases and then decreases asymptotically to $\text{S}_\text{{Py}}\underset{J_{{\text{ex}}}\rightarrow \infty}{\longrightarrow}73$~({emu/cm$^3$})$^2$. 
On the other hand, for the optic mode, the area of Fe-Co layer decreases from the value of Fe-Co single layer precession area, $\text{S}_\text{{Fe-Co}}$~=~410~({emu/cm$^3$})$^2$, corresponding to $J_{\text{ex}}=0$ to 0, while the Py one starts in 0, then reaches a maximum, and decreases to 0 like the Fe-Co mode. Both modes go to 0, because the optical mode disappears at a certain $J_{{\text{ex}}}$ in X-band, as shown in Fig.~\ref{Intensity}~(b).
Figs.~\ref{area} (a) and (b) show the behavior of the optical and acoustic modes respectively, along the hard axis of magnetization. In both graphics, continuous and dashed lines correspond to the contribution of the FeCo and Py layers, respectively. This behavior is predicted for all the samples, along both the easy and hard axes. Fig.~\ref{contrPy_all} shows the simulated contribution of the Py layer for different samples, with varying Fe-Co layer thicknesses and the same Py thickness as in the experiment. As the thickness of Fe-Co increases, the precession area of the Py layer also increases, allowing a greater spin current to be injected. In addition, the exchange corresponding to the maximum precession area of the Py layer is $J_{\text{ex}}$~=~0.11~erg/cm$^2$ for the hard axis and goes from 0.27~erg/cm$^2$ in the thinnest sample to 0.34~erg/cm$^2$ for thicker films for the easy axis of magnetization, as the Fe-Co thickness increases. Simulations show that along the hard axis, the $J_{\text{ex}}$ for the maximun area remains at the same position even at larger thicknesses of Fe-Co.


The inset of Fig.~\ref{contrPy_all} shows the Py precession area for the sample with a 100~nm Fe-Co thickness. The magnetization precession area increases, while the position of the maximum remains unchanged compared to the sample with a 25~nm Fe-Co thickness. These results indicate a strong correlation between the magnetization precession area, $J_{\text{ex}}$, and the thickness of the underlayer.

It is interesting to understand why the area precession peak is present. In order to propose a model that contains the main contributions to this effect, we focus on the torques acting on the Py layer. In particular, we propose to consider only the torques exerted by the microwave excitation and the Fe-Co layer precession via the exchange interaction. In this model, we do not include any contributions from magnetic anisotropy, with the aim to keep only the relevant terms that give account for the observed effect.
The torque per unit area in the Py layer can be expressed as $\tau=-\text{M}_{Py}\text{t}_{Py}\bf{\hat{\mu}} \times \bf{H_t}$ \cite{Gilbert}, where $\bf{\hat{\mu}}$ is the magnetic moment of Py layer, and $\bf{H_t}$ is the effective magnetic field acting on the Py layer that gives rise to the torque. With the assumption that the precession area has a quadratic dependence on the microwave excitation, it is reasonable to consider the squared torque per unit area, averaged over one oscillation period. After developing the equation and taking into account that the square torque depends on the area, we arrive at the following expression. 

\begin{align}
\langle \tau^2 \rangle & \approx
\text{M}_{\mathrm{Py}}^2 \text{t}_{\mathrm{Py}}^2 \frac{h^2}{4}
+ J_{\mathrm{ex}}^2 \big[
({m}_{\varphi_1}-{m}_{\varphi_2})^2 \nonumber \\
\qquad & +
({m}_{\theta_1}-{m}_{\theta_2})^2
\big]
\label{Torque}
\end{align}

\begin{figure}[h!]
    \centering
    \includegraphics[width=0.9\linewidth]{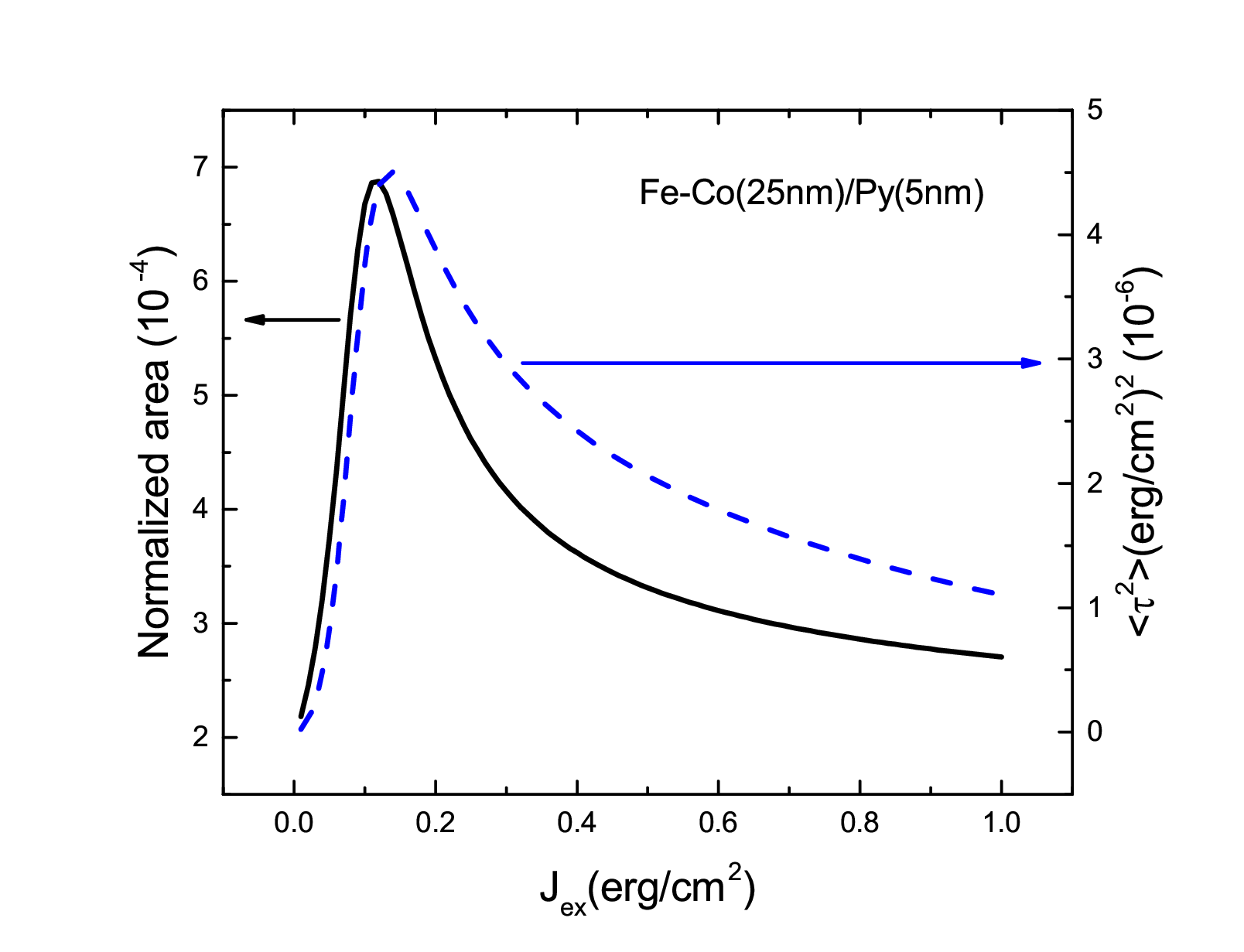}
    \caption{Normalized precession area (left axis) and average squared torque per unit area (right axis) in function of $J_{\text{ex}}$.}
    \label{torque}
\end{figure}

By analyzing Eq.~(\ref{Torque}) in the limiting cases, we observe that for $J_{\text{ex}}$ the equation depends only on the first term, which is associated with the Py layer and the microwave excitation. As $J_{\text{ex}}$ increases, the torque is expected to increase due to the quadratic dependence on the exchange coupling. On the other hand, in the limit $J_{\text{ex}}\rightarrow \infty$ the bilayer should behave as a single effective material, for that reason ${m}_{\varphi_1}={m}_{\varphi_2}$ and ${m}_{\theta_1}={m}_{\theta_2}$, and the torque-related term vanishes, since there is no longer an additional layer exerting torque on the other, but rather a single collective magnetic system. Therefore, it is reasonable to expect that the torque initially increases with $J_{\text{ex}}$, reaches a maximum, and then decreases. Plotting the behavior of $\langle\tau^2\rangle$ as a function of $J_{\text{ex}}$ we obtain the dependence shown in Fig.~\ref{torque},

The main approximations underlying the last model are that the intensity depends exclusively on $\langle\tau^2\rangle$, and that the contribution of magnetic anisotropy to the effective magnetic field responsible for the torque has been neglected; for these reasons, slight differences can be observed between these curves. The above presented simulations predict that the spin current injected from the Py layer in a coupled ferromagnetic bilayer can be maximized at an intermediate exchange coupling. This effect arises from the sum between microwave excitation and exchange-mediated torque between layers.

\begin{figure}[h!]
    \centering
    \includegraphics[width=1\linewidth]{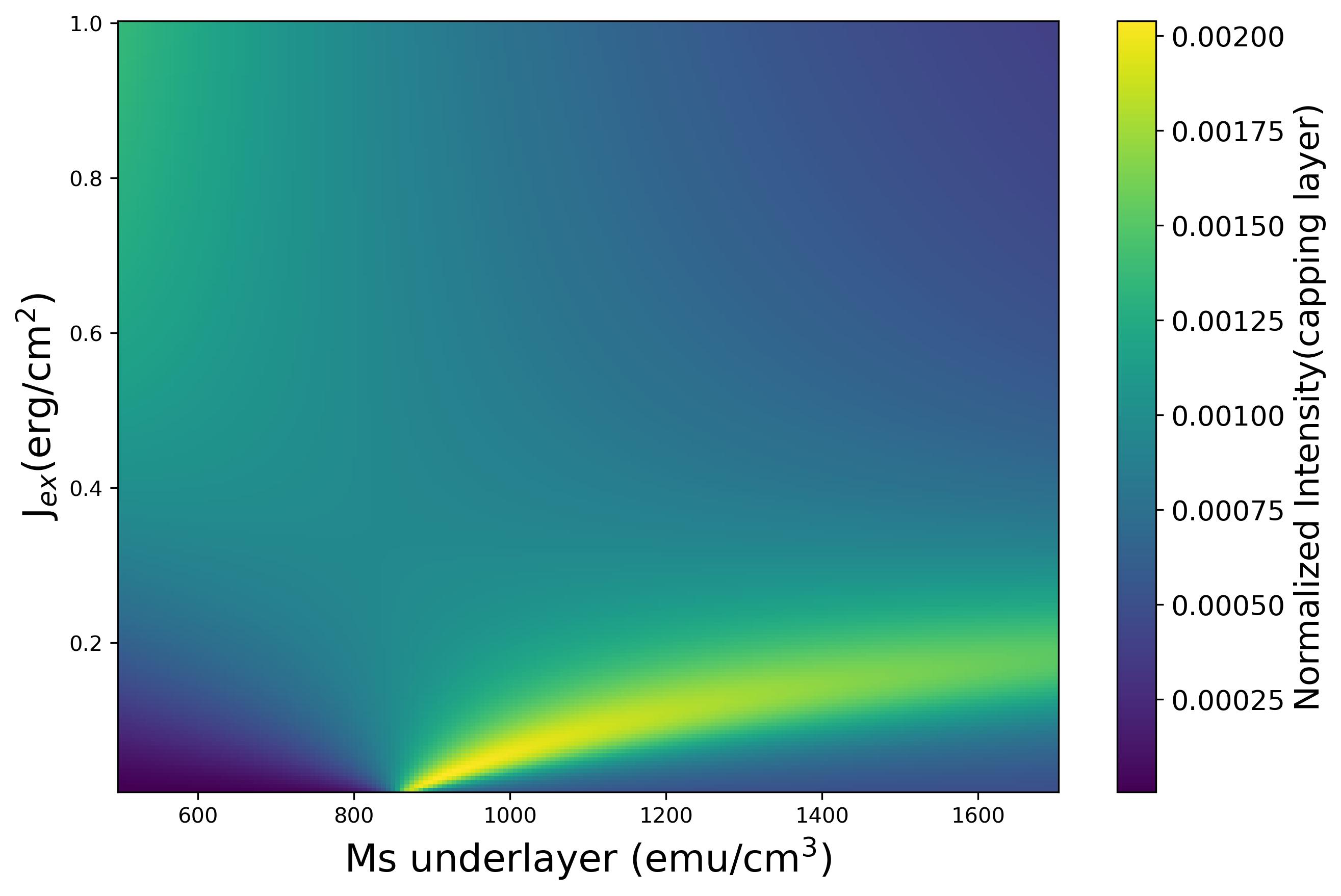}
    \caption{(a) Normalized precession area for the capping layer versus $J_{\text{ex}}$(y-axis) and Ms of the underlayer material (x-axis), for a capping material with $\text{M}_{\text{s}}$=480~emu/cm$^3$.}
    \label{colormap}
\end{figure}

In our experiments, $J_{\text{ex}} > 0.8$~erg/cm$^2$, however, in the X-band, the maximum precession area occurs around 0.11~erg/cm$^2$. If the frequency is increased, it is possible to shift the maximum toward higher $J_{\text{ex}}$ values; for example, it reaches 0.82~erg/cm$^2$ at 26~GHz. It is also possible to modify the maximum position and intensity by changing the $\text{M}_{{\text{s}}}$ of the underlayer and capping material. For fixed values for the $\text{M}_{{\text{s}}}$ of the capping layer, the low underlayer $\text{M}_{{\text{s}}}$ regime is dominated by the $J_{\text{ex}}$ term, leading to a nearly monotonic increase with increasing coupling strength. In this regime, the contribution associated to $({m}_{\varphi_1}-{m}_{\varphi_2})^2$ and $({m}_{\theta_1}-{m}_{\theta_2})^2$ term in Eq.~(\ref{Torque}) is negligible. This contribution becomes relevant only as the underlayer $\text{M}_{{\text{s}}}$ increases, which is precisely where the maximum observed emerges. Moreover, when the capping layer $\text{M}_{{\text{s}}}$ is varied while keeping the underlayer material fixed, lower values of Py $\text{M}_{{\text{s}}}$ enhances the torque that the bottom layer can exert on the top one. Therefore, for the capping layer, a material with very low $\text{M}_{{\text{s}}}$ would be preferable. Fig.~\ref{colormap} presents the normalized magnetization precession area of the capping material as a function of $J_{\text{ex}}$ and the underlayer $\text{M}_{{\text{s}}}$ with color indicating its magnitude.

This idea leads to the conclusion that we can use as underlayer a material with a $\text{M}_{{\text{s}}}$ larger than the capping material. In that way, for thicker underlayer and thin capping, the injected spin current in the external layer will be enhanced. In addition to this, by tuning the exchange constant to a value that maximizes the area of precession of magnetization, as shown in the Fig.~\ref{contrPy_all}, we could maximize the injected spin current. This could be achieved by placing a thin non-magnetic material between both ferromagnetic layers, and/or by selecting materials with $\text{M}_{{\text{s}}}$ that maximize the torque exerted on the capping layer. In that way, by maximizing the spin current injected, we can achieve a better efficiency in the spin injection, improving the energy efficiency, and achieving more efficient experiments for the generation of spin currents.

\section{Conclusions}

The dynamic magnetic properties of the bilayer system Fe\texorpdfstring{$_{85}$}{85}Co\texorpdfstring{$_{15}$}{15}/Py grown on MgO[100] was studied experimentally and theoretically. Using ferromagnetic resonance, the angular dependence of the resonance field was obtained, and it was fitted using a bilayer model, considering a system with uniaxial anisotropy, cubic magnetocrystalline anisotropy, and exchange constant $J_{\text{ex}}$. The existence of a hard and an easy magnetization axis along the Fe-Co~[110] and FeCo[100] directions was determined, respectively. It was observed that $\text{H}_r$ presents minima at $45^\circ$, which is the Fe-Co[100], coincident with the results obtained in the VSM and MOKE. The model quantitatively determines the magnetization precession as a function of $J_{\text{ex}}$. The simulations show that the precession area of the capping layer reaches a maximum at a certain value of $J_{\text{ex}}$ and increases with the underlayer thickness due to the torque exerted by the underlayer. These results indicate that the magnetization precession trajectory area, and consequently the spin current injection, can be maximized by tuning the exchange constant, the saturation magnetization of the materials involved and the excitation frequency.

\section*{Data Availability Statement}
The data that support the findings of this study are openly available in Zenodo at Ref.~\cite{Data_FeCoPy_2026}. The simulation code used to generate the numerical results is openly available in Zenodo at Ref.~\cite{Code_Eigenvectors_2026}.

\appendix
\section{Torque model}

To describe the torque exerted on the Py layer, we propose the following model, where the torque per unit area acting on the Py layer can then be expressed as
\begin{equation}
\boldsymbol{\tau} = -\frac{1}{A}{\vec{\boldsymbol{\mu}}_{p}\mathbf{H}_t} = -M_{\mathrm{Py}} t_{\mathrm{Py}}\,\hat{\boldsymbol{\mu}}_{p}\times \mathbf{H}_t.
\tag{A1}
\label{eq:A_torque}
\end{equation}
The effective magnetic field acting on the Py layer, $\mathbf{H}_t$, can be written as
\[
\mathbf{H}_t = \mathbf{H}_j+\mathbf{h}(t).
\]
Here,
\[
\mathbf{H}_j = -\frac{\partial E_j}{\partial \boldsymbol{\mu}_2}
= \frac{J}{M_2 t_2}\,\hat{\boldsymbol{\mu}}_1
\]
is the magnetic field in the Py layer arising from the exchange interaction,
where
\[
E_j = -J\,\hat{\boldsymbol{\mu}}_{1}\cdot\hat{\boldsymbol{\mu}}_{2}
     = -\frac{J}{M_2 t_2}\,\boldsymbol{\mu}_{2}\cdot\hat{\boldsymbol{\mu}}_{1},
\]
\[
\hat{\boldsymbol{\mu}}_{1} = \left(m_{{\theta}_1} \cos(\omega t),\, m_{{\varphi}_1} \sin(\omega t),\, 1\right),
\]
\[
\hat{\boldsymbol{\mu}}_{2} = \left(m_{{\theta}_2} \cos(\omega t),\, m_{{\varphi}_2} \sin(\omega t),\, 1\right),
\]
and the subscripts $1$ and $2$ refer to the Fe-Co and Py layers, respectively.
And $\mathbf{h}(t)$ is the microwave field, given by
\[
\mathbf{h}(t)=\left(-\sin(\omega t),\,\cos(\omega t),\,0\right).
\]

Solving \ref{eq:A_torque} and taking the square and obtaining the average value over a precession period, we obtain the following equation

\[
\begin{split}
\langle \tau^2 \rangle \approx\;&
M_{\mathrm{Py}}^2 t_{\mathrm{Py}}^2 \frac{h^2}{4} \\
&+ J_{\mathrm{ex}}^2 \Big[
\left(m_{\varphi_1}-m_{\varphi_2}\right)^2
+
\left(m_{\theta_1}-m_{\theta_2}\right)^2
\Big]
\end{split}
\]

\nocite{apsrev42Control}
\bibliographystyle{apsrev4-2}
\bibliography{bibliografia}

\end{document}